\documentclass[onecolumn,preprintnumbers,amsmath,amssymb,aps,prd,nofootinbib]{revtex4}

\usepackage{graphicx,epsfig}

\usepackage{bm}

\usepackage{amsmath}

\def\ii{{\rm i}}
\def\be{\begin{equation}}
\def\ee{\end{equation}}
\def\beq{\begin{eqnarray}}
\def\eeq{\end{eqnarray}}
\def\ie{{\it i.e.,}\ }
\def\eg{{\it e.g.,}\ }
\def\jmart{{JMaRT}\ }

\begin{document}

\centerline{}


\title{Instability of non-supersymmetric smooth geometries}

\author{Vitor Cardoso}
\email{vcardoso@phy.olemiss.edu} \affiliation{Department of
Physics and Astronomy, The University of Mississippi, University,
MS 38677-1848, USA \footnote{Also at: Centro de F\'{\i}sica
Computacional, Universidade de Coimbra, P-3004-516 Coimbra,
Portugal}}
 \author{\'Oscar J. C. Dias}
 \email{odias@perimeterinstitute.ca}
 \author{Jordan L. Hovdebo}
 \email{jlhovdeb@sciborg.uwaterloo.ca}
 \author{Robert C. Myers}
\email{rmyers@perimeterinstitute.ca} \affiliation{Perimeter
 Institute for Theoretical Physics,
 Waterloo, Ontario N2L 2Y5, Canada  \\
         and \\
 Department of Physics, University of Waterloo, Waterloo, Ontario
 N2L 3G1, Canada }


\begin{abstract}

Recently certain non-supersymmetric solutions of type IIb
supergravity were constructed \cite{ross}, which are everywhere
smooth, have no horizons and are thought to describe certain
non-BPS microstates of the D1-D5 system. We demonstrate that these
solutions are all classically unstable. The instability is a
generic feature of horizonless geometries with an ergoregion. We
consider the endpoint of this instability and argue that the
solutions decay to supersymmetric configurations. We also comment
on the implications of the ergoregion instability for Mathur's
`fuzzball' proposal.

\end{abstract}


\maketitle


\section{Introduction}


String theory has made great progress in understanding the
microphysics of black holes. In particular, for certain (nearly)
supersymmetric black holes, one is able to show that the
Bekenstein-Hawking entropy $S_{\rm BH}=A_{\rm hor}/4G$, as
computed in the strongly-coupled supergravity description, can be
reproduced in a weakly-coupled D-brane description as the
degeneracy of the relevant microstates \cite{2}
--- for reviews, see \cite{revall}. The AdS/CFT
correspondence \cite{big} provides further insights into these
issues by providing a dictionary relating the geometric description
of the physics in the near-horizon region with the physics of a dual
conformal field theory --- see \cite{adscft} for a review. In
particular, the AdS/CFT indicates that Hawking evaporation should be
a unitary process, in keeping with the basic tenets of quantum
theory. The discussion of black holes in the context of the AdS/CFT
correspondence makes evident that the path integral over geometries
in the bulk may include multiple saddle-points, \ie several
classical supergravity solutions, as found \eg in
\cite{thermal1,thermal2,fairy}.\footnote{Of course, in these
examples, a single saddle-point typically dominates the path
integral.} Another point that was realized early on is that the
geometric description of individual microstates would not have a
horizon \cite{pure,amati}.

In recent years, Mathur and collaborators have incorporated these
ideas in a radical revision of the stringy description of black
holes --- for a review, see \cite{fuzzy}. They argue that each of
the CFT microstates corresponds to a separate spacetime geometry
with no horizon. The black hole is dual to an ensemble of such
microstates and so the black hole geometry only emerges in a
coarse-grained description which `averages' over the $e^{S_{\rm
BH}}$ microstate geometries. In particular, this averaging should
produce an effective horizon at a radius where the individual
microstate geometries start to `differ appreciably' from one another
\cite{11,10}. Therefore in this scenario, quantum gravity effects
are not confined close to the black hole singularity, rather the
entire interior of the black hole is `filled' by fluctuating
geometries
--- hence this picture is often referred to as the `fuzzball'
description of black holes. The first support for this proposal came
from finding agreement between the propagation time of excitations
in the throat of certain microstate geometries and in the dual brane
description \cite{10,19}. A further remarkable feature, that has
drawn attention to these ideas, is that there is growing evidence
that the microstate geometries may be smooth, as well as
horizon-free.\footnote{`Smooth' means the curvature is finite
everywhere up to orbifold singularities. The curvatures in the
throat may also be very large.} In the case of the D1-D5 system,
smooth asymptotically flat geometries can be constructed
corresponding to all of the RR ground states in the dual CFT
\cite{10,two}. Despite their large degeneracy, this two-charge
system will not produce a macroscopic black hole horizon. However, a
large horizon can be produced by introducing a third charge,
Kaluza-Klein momentum \cite{early,early1}. Recently progress has
been made in constructing smooth microstate geometries in the
D1-D5-P system \cite{three1,three12,three2}. While large families of
such solitons are now known, a complete understanding of the
three-charge case remains to be found. Further preliminary work on
the four charge system of D1-D5-P-KK has also appeared \cite{four}.

In general, the preceding discussion connecting microstates with
smooth geometries focuses on supersymmetric configurations. This
raises the interesting question of how the fuzzball proposal would
be extended to non-supersymmetric black holes. In particular, are
there non-supersymmetric versions of the smooth horizon-free
geometries corresponding to non-BPS microstates? Remarkably,
Jejjala, Madden, Ross and Titchener \cite{ross} recently extended
the known set of D1-D5 microstate geometries with a family of
non-supersymmetric solutions, hereafter referred to as \jmart
solitons. The \jmart solutions comprise a five-parameter family of
non-supersymmetric smooth geometries which are asymptotically
flat.\footnote{By considering orbifolding, this family can extended
by a third integer \cite{ross} but we will focus on the original
five-parameter solutions.} These solutions may be parameterized by
the D1-brane and D5-brane charges, the (asymptotic) radius of the
internal circle with Kaluza-Klein momentum, and by two integers $m$
and $n$ which fix the remaining physical parameters. These integers
also determine a spectral flow in the CFT which allows the
underlying microstate to be identified. For $m=n+1$, the \jmart
solitons reduce to supersymmetric solutions found previously in
\cite{two,three1,three12}.

An important feature which distinguishes the \jmart solitons from
any of the analogous supersymmetric solutions is the presence of
an ergoregion. As a consequence, in these non-supersymmetric
geometries, there is an inner region (that extends to the origin)
where states of negative energy are allowed. This then leads
naturally to the question of whether or not the ergoregion
produces an instability of the  background. One possibility is
that the ergoregion may lead to superradiant scattering which can
produce a catastrophic instability in some situations
\cite{press,dnr,bhb}. However in the present case, this
possibility is easily dismissed \cite{ross} because the solutions
are horizon-free. Since the seminal work of Zel'dovich \cite{zel}
on superradiant amplification of electromagnetic waves incident upon
an absorbing cylinder, it has been known that the key
ingredients for superradiance is the existence of an ergoregion
{\it and} an absorbing surface. For black holes, the horizon plays
the latter role but certainly the \jmart geometries lack such a
surface.

Quite interestingly, there is another class of instabilities,
which we simply refer to as `ergoregion instabilities', that
generically afflict spacetime geometries with an ergoregion, but
no horizon. These instabilities were first discovered by Friedman
\cite{friedman}, who provided a very general discussion. Explicit
computations of the instability were later made in
\cite{cominsschutz,compute} for the case of rotating stars with an
ergoregion. There the existence of this instability was explicitly
verified for a free scalar field in the background of a rotating
star. According to Friedman's general arguments however, the
instability should also exist for electromagnetic and
gravitational waves. Since the \jmart solutions \cite{ross} have
an ergoregion but no horizon, one might suspect that a similar
ergoregion instability would arise in these geometries. The
present paper then explicitly verifies the presence of an
ergoregion instability for the \jmart backgrounds with a variety
of techniques. Further we consider the endpoint of the resulting
decay and argue that it should be a smooth supersymmetric
solution.

Our results have immediate consequences for the endpoint of
tachyon decay discussed in \cite{rossTC}. There, Ross extended the
discussion of \cite{horowitzTC} to D1-D5 black strings for which
he identified tachyonic string modes in a particular winding
sector. He argued that the condensation of these tachyons would
transform the spacetime to a \jmart soliton. In conjunction with
the above results, we see that these solutions cannot be the final
endpoint of these decays but rather they should end with a
supersymmetric microstate geometry. Our analysis and the
ergoregion instability may also have interesting implications for
Mathur's fuzzball proposal more generally.

The remainder of our paper is organized as follows: Section
\ref{general} provides a brief exposition on Friedman's analysis
\cite{friedman}. In Section \ref{formalism}, we briefly review
some of the features of the \jmart solutions and present the main
equations used in the subsequent analysis, namely the radial and
angular wave equations for a free massless scalar field, as well
as some of their properties. In Section \ref{wkb} we compute the
details of the instability using a WKB approach
\cite{cominsschutz}. We show explicitly that the instability
exists for a general non-supersymmetric geometry of \cite{ross},
and that it disappears for supersymmetric objects, as expected. In
Section \ref{sec:Match}, we use an alternative method, that of
matched asymptotic expansions, to investigate the instability and
its properties. The methods of sections \ref{wkb} and
\ref{sec:Match} are complementary, \ie their regime of validity is
different. We then perform a numerical analysis of the wave
equation in Section \ref{numerical} to complement the analytical
calculations. We find that the results of both analytical analyses
agree remarkably well with the numerical results. In section
\ref{conclusion}, after summarizing the main properties of the
ergoregion instability, we discuss various related topics: the
endpoint of this instability; its consequences for Ross's tachyon
condensation \cite{rossTC}; general implications for the fuzzball
picture of black holes.


\section{\label{general}{Ergoregion instabilities}}


There are two classes of instabilities that are of potential
interest for the \jmart backgrounds \cite{ross} (or
non-supersymmetric geometries in general), namely: the superradiant
instability, and the ergoregion instability. In this section, we
demonstrate why superradiance is not present in these geometries,
as first noted in \cite{ross}, and we introduce the general
argument of \cite{friedman} that suggests an ergoregion
instability is present. In the following sections, we verify the
presence of the ergoregion instability with a complete analytic
and numerical analysis of its properties.


\subsection{Geometries with an ergoregion and horizon:
Superradiance}


For a general (stationary asymptotically flat) black hole, the
equations describing spin-$s$ fields can always be written as
\be \frac{d^2\Psi}{dr_*^2}+V(\omega,\,r)\Psi=0\,\,\label{wave}\ee
where $\omega$ was introduced with a Fourier transform with
respect to the asymptotic time coordinate: $\Psi(t)=e^{-i\omega
t}\Psi(\omega)$. The radius $r_*$ is a convenient tortoise
coordinate and in general one finds:
\begin{equation}
\left\{
\begin{array}{ll}
r_* \sim r \,,&   V \sim \omega ^2\,\,\, \,\,\,\,\quad
\quad {\rm as}\ r\rightarrow \infty \,, \\
e^{r_*} \sim (r-r_+)^{\alpha} \,,&   V \sim (\omega-\Phi)^2\,\,\,
{\rm as}\ r\rightarrow r_+ \,,
\end{array}
\right. \label{bound}
\end{equation}
where $\alpha$ is a positive constant. The potential $\Phi$ can be
a rotational potential (in the Kerr geometry $\Phi=m\Omega$, with
$m$ an azimuthal number, and $\Omega$ the angular velocity at the
horizon) or a chemical potential (in the Reissner-N$\ddot{\rm
o}$rdstrom geometry, $\Phi=qQ$, where $q$ is the charge of the
field and $Q$ the charge of the black hole).

For a wave scattering in this geometry, Eq.~(\ref{wave}) yields
the following asymptotic behavior:\footnote{Implicitly, we
consider a massless field here but the discussion is generalized
to massive fields in a straightforward way.}
\begin{equation}
\Psi _1 \sim\left\{
\begin{array}{ll}
\mathcal{T}\,(r-r_+)^{-\ii \alpha (\omega-\Phi)} & {\rm as}\
r\rightarrow r_+ \,,\label{bound2} \\
\mathcal{R}\,{\rm e}^{\ii {\omega r}}+ {\rm e}^{-\ii {\omega r}}&
{\rm as}\ r\rightarrow \infty\,.
\end{array}
\right.
\end{equation}
These boundary conditions correspond to an incident wave of unit
amplitude from $+\infty$ giving rise to a reflected wave of
amplitude $\mathcal{R}$ going back to $+\infty$ and a transmitted
wave of amplitude $\mathcal{T}$ at the horizon --- the boundary
condition introduces only ingoing waves at the horizon. Assuming a
real potential (which is almost always the case) the complex
conjugate of the solution $\Psi _1$ satisfying the boundary
conditions (\ref{bound2}) will satisfy the complex-conjugate
boundary conditions:
\begin{equation} \Psi _2
\sim\left\{
\begin{array}{ll}
\mathcal{T}^*(r-r_+)^{\ii \alpha (\omega-\Phi)} & {\rm as}\ r\rightarrow r_+ \,, \\
\mathcal{R}^*{\rm e}^{-\ii {\omega r}}+ {\rm e}^{\ii {\omega r}}&
{\rm as}\ r\rightarrow \infty\,.
\end{array}
\right. \label{bound3}
\end{equation}
Now, these two solutions are linearly independent, and the standard
theory of ODE's tells us that their Wronskian, $W=\Psi _1
\partial_{r_*}\Psi _2 -\Psi _2
\partial_{r_*}\Psi _1$, is a constant (independent of $r$). If we evaluate
the Wronskian near the horizon, we get $W=
-2\ii(\omega-\Phi)|\mathcal{T}|^2$, and near infinity we find
$W=2\ii \omega(|\mathcal{R}|^2-1)$. Equating the two we get
\begin{equation}
 |\mathcal{R}|^2=1-\frac{\omega -\Phi}{\omega}|\mathcal{T}|^2\,.
\end{equation}
Now, in general $|\mathcal{R}|^2$ is less than unity, as is to be
expected. However, for $\omega-\Phi<0$ we have that
$|\mathcal{R}|^2>1$. Such a scattering process, where the reflected
wave has actually been amplified, is known as superradiance. Of
course the excess energy in the reflected wave must come from that
of the black hole, which therefore decreases.

Superradiant scattering can lead to an instability if, \eg we have
a reflecting wall surrounding the black hole that scatters the
returning wave back toward the horizon.  In such a situation, the
wave will bounce back and forth, between the mirror and the black
hole, amplifying itself each time. The total extracted energy
grows exponentially until finally the radiation pressure destroys
the mirror.  This is Press and Teukolsky's black hole bomb, first
proposed in \cite{press}. This instability can arise with an
effective `mirror' in a variety of situations: a scalar field with
mass $\mu> \omega$ in a Kerr background creates a potential that
can cause flux to scatter back toward the horizon
\cite{detweiler}; infinity in asymptotically AdS spaces also
provides a natural wall \cite{bhbAdS} that leads, for certain
conditions, to an instability; a wave propagating around rotating
black branes or rotating black strings may similarly find itself
trapped \cite{cardoso}.


\subsection{\label{free}Geometries with ergoregion but horizon-free:
Ergoregion instability}


Suppose now there is no horizon in the background spacetime. What
changes with respect to the former discussion is the boundary
conditions: since there is no horizon and no absorption. For this
case, the boundary condition (\ref{bound2}) at the horizon is
replaced by some kind of regularity condition at the origin. We
suppose the radial coordinate $r$ now ranges from zero to infinity
and we impose the following boundary condition:
\begin{equation}
\Psi \sim A f(r)\,\,,\,\,\,r\rightarrow 0\,,
\end{equation}
where $f(r)$ is some well-behaved {\it real} function. This ansatz
encompasses for instance typical regularity requirements where,
\eg one chooses $f(r) \sim r^{\beta}$ with $\beta>0$. Repeating
the above calculation, one gets $|\mathcal{R}|^2=1$. Therefore the
absence of a horizon, which precludes any absorption, prevents
superradiance and hence the superradiant instability.

Nevertheless, geometries with an ergoregion but without horizons
are the arena of another class of instability. This ergoregion
instability was discovered by Friedman \cite{friedman}. Even
though his discussion was made in four-dimensions only, it is
trivial to extend it to any number of dimensions. The instability
arises because of the following \cite{friedman}: Given the test
field energy-momentum tensor $T^{ab}$, we can associate a
canonical energy
\begin{equation}
 {\cal E}_S=\int_S t^a\, T_a{}^b  dS_b\,,
\label{canon}
\end{equation}
where $t^{a}$ is the background Killing vector which generates
time translations in the asymptotic geometry. Now, because $t^a$
is space-like within an ergosphere, initial data can be chosen on
a Cauchy surface S which makes ${\cal E}_S$ negative. Moreover, it
is shown in \cite{friedman} that the energy can be negative only
when the test field is time dependent. Then, since the field is
time dependent, and since only positive energy can be radiated at
future null infinity, the value of ${\cal E}_S$ can only decrease
further from one asymptotically null hypersurface $S$ to another,
say, $S'$, in the future of $S$. Thus, the energy ${\cal E}_S$
will typically grow negative without bound. This instability was
computed analytically using a WKB approximation in
\cite{cominsschutz} for rotating stars. There it was shown that
the instability timescale is usually very large (typically larger
than the age of the universe). The analysis of \cite{cominsschutz}
was improved in \cite{compute} where further details of the
instability were computed numerically.

A key assumption above is that the system can not settle down to a
negative energy configuration which, while time dependent, is
nonradiative. Friedman \cite{friedman} was able to rule out such
marginal cases where ${\cal E}_S$ is negative but constant for a
four-dimensional massless scalar or electromagnetic fields.
However, in fact, one is able to identify negative energy bound
states for the \jmart backgrounds --- see Appendix \ref{sec:A2}
--- and so a more thorough analysis is called for. Hence, in the following,
we apply a variety of techniques to explicitly show that these
microstate geometries suffer from an ergoregion instability.

\section{\label{formalism} Formalism}

We now consider wave propagation of a free massless scalar field
in the \jmart backgrounds \cite{ross}, and from this identify an
ergoregion instability for these geometries in the subsequent
sections. The \jmart solutions are described in detail in
\cite{ross} and are quite involved. We will provide a brief
discussion of some of the properties of these solutions here, but
will refer the reader to \cite{ross} for the full details.

The \jmart solitons are solutions of type IIb supergravity
corresponding to three-charge microstate geometries of the D1-D5-P
system. The system is compactified to five dimensions on $T^4 \times
S^1$ with the D5-branes wrapping the full internal space and the
D1-branes and KK-momentum on the distinguished $S^1$. The notation
is best understood by considering the construction of these
solutions. One begins with the general solutions of \cite{early1}
which contain eight parameters: a mass parameter, $M$; spin
parameters in two orthogonal planes, $a_1,a_2$; three boost
parameters, $\delta_1,\delta_5,\delta_p$, which fix the D1-brane,
D5-brane and KK-momentum charges, respectively; the radius of the
$S^1$, $R$; the volume of the $T^4$ (which plays no role in the
following). The geometry is described by the six-dimensional line
element written in Eq. (2.12) of \cite{ross}, which is parameterized
by a time coordinate $t$; a radial coordinate $r$; three angular
coordinates $\theta$, $\phi$, $\psi$; and the coordinate on the
$S^1$, $y$.

One then imposes a series of constraints to ensure that the
solutions are free of singularities, horizons and closed time-like
curves. In particular, one focuses on a low-mass regime,
$M^2<(a_1-a_2)^2$, in which no black holes exist. Then one finds
solitonic solutions where an appropriate circle shrinks to zero at
the origin and the constraints ensure that this happens smoothly.
First, $M$ and $R$ can be fixed in terms of the remaining
parameters
--- see Eqs. (3.15) and (3.20) of \cite{ross}. Two quantization
conditions constrain the remaining parameters in terms of two
integers $m,n$ \cite{ross}:
\begin{eqnarray}
\frac{j+j^{-1}}{s+s^{-1}}=m-n\,, \qquad
\frac{j-j^{-1}}{s-s^{-1}}=m+n\,,
 \label{ross 1a}
\end{eqnarray}
where $j=\sqrt{\frac{a_2}{a_1}}\leq 1$ and $s=\sqrt{\frac{s_1 s_5
s_p}{c_1 c_5 c_p}}\leq 1$. We are using the notation here that
$c_i \equiv \cosh \delta_i$ and $s_i \equiv \sinh \delta_i$.
Without loss of generality, one assumes $a_1 \ge a_2\ge 0$ which
implies $m>n\ge0$. We also note here that the special case $m=n+1$
corresponds to supersymmetric solutions.

This leaves a five-parameter family of smooth solitonic solutions.
We can think of the independent parameters as the D1-brane and
D5-brane charges, $Q_1,Q_5$; the (asymptotic) radius of the
$y$-circle, $R$; and the two integers, $m$ and $n$, which fix the
remaining physical parameters as \cite{ross}
\begin{equation}
Q_P=nm\frac{Q_1Q_5}{R^2}\,,\quad J_\phi=-m\frac{Q_1Q_5}{R}\,,\quad
J_\psi=n\frac{Q_1Q_5}{R}\,. \label{simple}
\end{equation}
Of course, depending on the specific application, it may be more
appropriate and/or simpler to describe the solutions using a
different set of quantities. In our case, when we make explicit
calculations of the ergoregion instability, we will fix the
parameters $n,m,a_1,c_1$ and $c_5$ or $c_p$. As we are interested
in non-supersymmetric backgrounds, we also impose $m\geq n+2$. To
conclude our discussion of notation, we add that the roots of
$g^{rr}$, $r_+$ and $r_-$, will also appear in the following but
they are determined by $M$ and the spin parameters
--- see Eq.~(3.2) of \cite{ross}.

The key ingredient producing the instability in the \jmart
solutions is the existence of an ergoregion. To verify the
presence of the ergoregion, one takes as usual the norm of the
Killing vector $V=\partial_t$ and using Eq.~(2.12) of \cite{ross},
calculates
\begin{eqnarray}
 g_{\mu\nu}V^{\mu}V^{\nu} =-\frac{f-Mc_p^2}{\sqrt{\tilde{H}_1
 \tilde{H}_5}}\,,
 \label{ergoregionV}
\end{eqnarray}
where $f(r)=r^2+a_1^2\sin^2\theta+a_2^2\cos^2\theta >0$ and
$\tilde{H}_i(r)=f(r)+M s_i^2$, $i=1,5$. It is then clear that
$V=\partial_t$ becomes space-like for $f(r)<M$ and thus an
ergosphere appears at $f(r)=M$. An inspection of the metric also
allows one to conclude the geometry rotates along $\phi$, $\psi$
and $y$ since $g_{t\phi}\neq 0$, $g_{t\psi}\neq 0$ and $g_{t
y}\neq 0$. The supersymmetric limit of the \jmart solitons
corresponds to take the limit $M\rightarrow 0$ and $\delta_i
\rightarrow \infty$, while keeping the other parameters fixed,
including the conserved charges $Q_i=M s_i c_i$ \cite{ross}. So,
in the supersymmetric limit the norm becomes $|V|^2=-f /
\sqrt{\tilde{H}_1 \tilde{H}_5}$, which is always negative and thus
the ergoregion is not present.

Now consider the Klein-Gordon equation for a massless scalar field
propagating in the \jmart geometries,
\begin{equation} \frac{1}{\sqrt{-g}}\frac{\partial}{\partial x^{\mu}}
\left(\sqrt{-g}\,g^{\mu \nu}\frac{\partial}{\partial x^{\nu}}\Psi
\right)=0\,. \label{klein}
\end{equation}
Implicitly, we are using the string-frame metric in which case one
can think of eq.~(\ref{klein}) as the linearized equation of
motion for the Ramond-Ramond scalar. As described above, these
backgrounds can be thought of as special cases of the general
D1-D5-P solutions found earlier \cite{early1} and so one may apply
separation of variables following \cite{finnx}. Introducing the
following ansatz \footnote{Note that the negative sign for $\lambda$ corrects
a typo found in \cite{ross}}
\begin{eqnarray}
\Psi=\exp\left[-i\omega \frac{t}{R}-i\lambda \frac{y}{R}+i
m_{\psi} \psi +i m_{\phi} \phi \right]\, \chi(\theta)\,h(x) \,,
 \label{separation ansatz}
\end{eqnarray}
one gets an angular equation
\begin{eqnarray}
 \frac{1}{\sin{2\theta}}\frac{d}{d\theta}\left
(\sin{2\theta}\,\frac{d\chi}{d\theta}\right )+{\biggr
[}\Lambda-\frac{m_{\psi}^2}{\cos^2\theta}-\frac{m_{\phi}^2}{\sin
^2\theta}+\frac{\omega ^2-\lambda ^2}{R^2}(a_1 ^2\sin
^2\theta+a_2^2\cos^2\theta) {\biggr ]}\chi=0 \label{angeq}\, ,
\end{eqnarray}
  and a radial equation\footnote{Note the factor $(r_+^2-r_-^2)$ that appears in the
two last terms of the lhs of (\ref{radialeq-r}), which are
necessary for dimensional consistency, corrects
the typo appearing in Eq. (6.4) of \cite{ross} }
  \begin{eqnarray}
\begin{aligned} & \frac{1}{r} \frac{d}{dr} \left[ \frac{g(r)}{r}
\frac{d}{dr} h \right]
  - \Lambda h + \left[ \frac{(\omega^2 - \lambda^2)}{R^2} (r^2 + M s_1^2 + M
    s_5^2) + (\omega c_p + \lambda s_p)^2 \frac{M}{R^2} \right] h
  \\
  & -(r_+^2-r_-^2)\frac{(\lambda - n m_\psi + m
  m_\phi)^2}{(r^2-r_+^2)} \, h +
  (r_+^2-r_-^2)\frac{(\omega \varrho + \lambda \vartheta - n m_\phi + m
    m_\psi)^2}{(r^2- r_-^2)} \, h = 0 \label{radialeq-r}\,,
\end{aligned}
\end{eqnarray}
where $g(r)=(r^2-r_+^2)(r^2-r_-^2)$, and we used
$\sqrt{-g}=r\sin\theta\cos\theta \sqrt{{\tilde H_1}{\tilde H_5}}$
(this is the determinant of the metric (2.12) of \cite{ross}). If we introduce a
dimensionless variable
\begin{equation}
x = \frac{r^2 - r_+^2}{r_+^2 - r_-^2},
\end{equation}
 we can rewrite the radial equation in the form
\begin{equation}
\partial_x[ x(x+1) \partial_x h]+ \frac{1}{4}
\left [ \kappa^{2}x + 1-\nu^2+
\frac{\xi^2}{x+1}-\frac{\zeta^2}{x}\right ] h =0 \,,
 \label{rad eq0}
\end{equation}
with
\begin{eqnarray}
& &
\kappa^{2}=(\omega^2-\lambda^2)\frac{r_+^2-r_-^2}{R^2}\,, \nonumber \\
& & \xi=\omega\varrho+\lambda \vartheta-m_{\phi}n+ m_{\psi} m\,, \nonumber \\
& & \zeta=\lambda - m_{\psi}n +m_{\phi} m \,,\nonumber \\
& & \varrho=\frac{c_1^2 c_5^2 c_p^2-s_1^2 s_5^2 s_p^2}{s_1 c_1 s_5
c_5} \,,\nonumber \\
& & \vartheta =\frac{c_1^2 c_5^2 -s_1^2 s_5^2}{s_1 c_1 s_5
c_5}\,s_p c_p\,,
 \label{rad eq parameters0}
\end{eqnarray}
and
\begin{equation}
\nu^2=1+\Lambda-\frac{\omega^2-\lambda^2}{R^2}(r_+^2+Ms_1^2+Ms_5^2)-(\omega
c_p+\lambda s_p)^2 \frac{M}{R^2}\,.
 \label{nu0}
\end{equation}
The quantities $\omega, \,\lambda, \, m_{\psi},\, m_{\phi}$ are
all dimensionless --- the last three being integers. Again, we
refer the reader to \cite{ross} for a detailed account of the
quantities appearing above. The reader should take note that our
notation is not in complete accord with that of \cite{ross}. That
is, to simplify our formulae in the following, we have defined
$\kappa\equiv 1/\sigma$, the inverse of the quantity $\sigma$ used
there.

Of critical importance in characterizing the solutions of the
scalar wave equation is the sign of $\kappa^{2}$.  The term $x
\kappa^2$ dominates at large $x$, determining the asymptotic
behavior of the solution.  In this paper we will mainly be
interested in outgoing modes so we choose $\kappa^{2}$ to be
positive.  The two remaining possibilities: $\kappa^{2}=0$ and
$\kappa^{2} < 0$, will be considered in the appendices.

The angular equation (\ref{angeq}) (plus regularity requirements) is
a Sturm-Liouville problem. We can label the corresponding
eigenvalues $\Lambda$ with an index $l$,
$\Lambda(\omega)=\Lambda_{lm} (\omega)$ and therefore the
wavefunctions form a complete set over the integer $l$. In the
general case, the problem at hand consists of  two coupled second
order differential equations: given some boundary conditions, one
has to compute {\it simultaneously} both values of $\omega$ and
$\Lambda$ that satisfy these boundary conditions. However, for
vanishing $a_i ^2$ we get the (five-dimensional) flat space result,
$\Lambda=l(l+2)$, and the associated angular functions are given by
Jacobi polynomials. For non-zero, but small $\frac{\omega ^2-\lambda
^2}{R^2}a_i ^2$ we have
\begin{eqnarray}
 \Lambda=l(l+2)+\mathcal{O}\left (a_i^2\frac{\omega^2-\lambda
^2}{R^2}\right) \label{app} \,.
\end{eqnarray}
The integer $l$ is constrained to be $l\geq
|m_{\psi}|+|m_{\phi}|$. We will always assume
$a_i^2\frac{\omega^2-\lambda ^2}{R^2} \ll {\rm
max}(m_{\psi}^2,m_{\phi}^2)$ (with $i=1,2$), thus $\Lambda \simeq
l(l+2)$. Making this assumption implies we may neglect the terms
proportional to $a_i$ in the angular equation, but given the way
$\Lambda$ and $\omega$ appear in the radial equation, the
corrections to $\Lambda$ may not be negligible when we determine
$\omega$.  To ensure that fixing $\Lambda=l(l+2)$ is consistent in
both the angular and radial equations we must additionally require
\begin{equation}
a_i^2 \ll \max \left ( | r_+^2+M(s_1^2+s_5^2) | , M c_p^2 \right ) \ ,
\end{equation}
so that the contribution to $\nu$ from the $a_i$ dependent
corrections of $\Lambda$ are negligible (see (\ref{nu0})).

Taking the complex conjugate of Eq.~(\ref{angeq}) we can see that
the exact solution to the angular equation has the symmetry
\be \Lambda_{lm}(-\omega^*)=\Lambda^*_{lm}(\omega)\,.\ee
With this symmetry, one can also check the following:
\beq & & (\nu^{2})^*(\omega,\lambda)=\nu^2(-\omega ^*,-\lambda)\,, \label{symmetry} \\
& &(\xi^{2})^*(\omega,\lambda,m_{\psi},m_{\phi})=\xi^2(-\omega
^*,-\lambda,-m_{\psi},-m_{\phi})\,,\\
& &(\zeta ^2)^*(\lambda,m_{\psi},m_{\phi})=\zeta
^2(-\lambda,-m_{\psi},-m_{\phi}) \,.\eeq
Therefore, from the wave equation (\ref{rad eq0}) it follows that
if $\omega$ is an eigenvalue for given values of
$m_{\psi},m_{\phi},\lambda$ with eigenfunction $h$, then $-\omega
^*$ is an eigenvalue for $-m_{\psi},-m_{\phi},-\lambda$ with
eigenfunction $h^*$. Furthermore, if $he^{-i\omega t}$ is outgoing
unstable, so is $h^*e^{i\omega ^*t}$. Since the symmetry
simultaneously flips all the signs of $m_{\psi},m_{\phi},\lambda$,
without loss of generality, we can only fix the sign of one, \eg
${\mathcal Re}( \omega ) \leq 0$.

To conclude this section, we point out that the angular equation
(\ref{angeq}) can be recast in the somewhat more familiar form:
\be
 \frac{1}{\sin\theta\cos\theta}\frac{d}{d\theta}\left
(\sin\theta\cos\theta\,\frac{d\chi}{d\theta}\right )+{\biggr
[}\hat{\Lambda}-\frac{m_{\psi}^2}{\cos^2\theta}-\frac{m_{\phi}^2}{\sin
^2\theta}+\frac{\omega ^2-\lambda ^2}{R^2}(a_2
^2-a_1^2)\cos^2\theta {\biggr ]}\chi=0 \label{angeqswsh}\,,
 \ee
where
\be \hat{\Lambda}=\Lambda+\frac{\omega ^2-\lambda^2}{R^2}a_1^2
\,.\ee
This is just the equation for a five-dimensional scalar spheroidal
harmonic \cite{teukolskyswsh} which arises, \eg in the separation
of Klein-Gordon equation in the background of a five-dimensional
rotating black hole \cite{myersperry}.

\section{\label{wkb} WKB analysis}

We now explicitly show that the \jmart geometries \cite{ross}
suffer from an ergoregion instability. As described above, this
instability is due to the fact that the geometry has an ergoregion
but no horizon. We shall identify modes of the scalar field that
are regular at the origin, represent outgoing waves at infinity
and grow with time. In this section, we follow the WKB analysis of
\cite{cominsschutz} and show that it applies to the
non-supersymmetric \jmart solutions, with the same qualitative
conclusions.

To begin, we want to write the radial wave equation in the form of
an effective Schr$\ddot{\rm o}$dinger equation. In order to do so,
we first transform to a new `wavefunction' $H$ defined with
\begin{eqnarray}
 h(x)=\frac{1}{\sqrt{x(1+x)}}\,H(x)\,.
 \end{eqnarray}
Inserting this in (\ref{rad eq0}), we get
\begin{eqnarray}
 -\partial ^2 _x H + U_{\rm eff}\,H=0 \label{Schrod 1}\,,
\end{eqnarray}
where
\begin{eqnarray}
 U_{\rm eff} = -\frac{ \kappa^{2}x^3 + (1-\nu ^2+
\kappa^{2})x^2 +(1-\nu^2+\xi^2-\zeta^2)x
+1-\zeta^2}{4x^2(1+x)^2}\,. \label{rad eq01}
\end{eqnarray}

Now in order to simplify our analysis, we choose: $\lambda=0$,
$m_{\phi}=0$, and large $m_{\psi}$. With $\lambda\ne0$, the waves
see a constant potential at infinity and thus the amplitude of the
outgoing waves can be suppressed there. We also consider
$l=m_{\psi}$ modes, which are expected to be the most unstable.
Modes with $l \gg m_{\psi}$ must be similar to modes with
$m_{\psi}=0$ for some $l$ and these are not unstable.  With these
choices, we have
\begin{eqnarray}
 \kappa^{2}&=&\omega ^2 \frac{r_+ ^2-r_-^2}{R^2}\,,\quad  \zeta
^2 = n^2m_{\psi}^2\,, \quad \xi ^2 = m^2 m_{\psi}^2+\omega ^2\varrho
^2+2\omega \varrho mm_{\psi}\,,
\\
1-\nu ^2 &\simeq & - m_{\psi}^2+\omega
^2\frac{r_+^2+Ms_1^2+Ms_5^2+Mc_p^2}{R^2} \,.
 \end{eqnarray}
Instead of working directly with the frequency of the wave, it
will be convenient to work with the pattern speed along the $\psi$
direction, which is the angular velocity at which surfaces of
constant phase rotate. This velocity is proportional to
\begin{eqnarray}
\Sigma_{\psi} = \frac{\omega}{m_{\psi}}\,,
 \label{Sigma}
\end{eqnarray}
where the proportionality constant $R^{-1}$ is always positive. It
is important to compare the sign of the pattern speed along $\psi$
with the sign of the angular velocity of the geometry along $\psi$
defined as usual by\footnote{Note that the geometry rotates
simultaneously along the $\psi$, $\phi$ and $y$ directions. We
find $\Omega_{\psi}$ using of (2.1), (3.17) and (3.19) of
\cite{ross}.}
\begin{eqnarray}
 \Omega_{\psi}=-\frac{g_{t\psi}}{g_{\psi\psi}}&=&
-\frac{2M s_p\, c_p \,\cos^2\theta}{\sqrt{\tilde{H}_1
\tilde{H}_5}}\frac{ R /n}{g_{\psi\psi}} \nonumber \\
&=& -\frac{2 Q_p \,\cos^2\theta}{\sqrt{\tilde{H}_1
\tilde{H}_5}}\frac{\cos^2\theta R /n}{g_{\psi\psi}} < 0, \quad
\forall \, x>0 \,,
 \label{Omega}
\end{eqnarray}
where $Q_p=M s_p\, c_p$ is the Kaluza-Klein momentum charge. So,
when $\Sigma_{\psi}$ is negative, the wave is propagating in the same
sense as the geometry.

\begin{figure}[h]
\begin{center}
\resizebox{8cm}{6cm}{\includegraphics{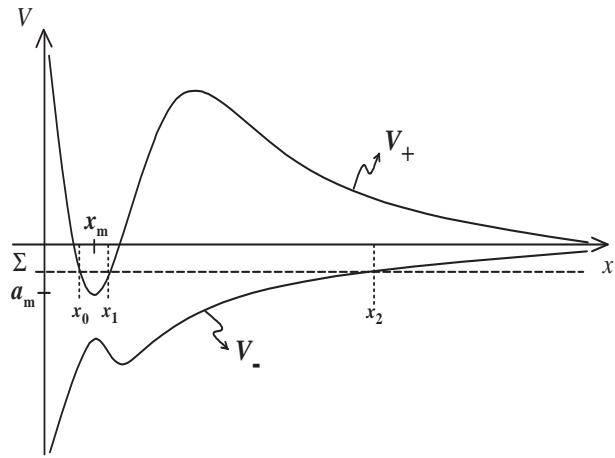}}
\end{center}
\caption{Qualitative shape of the potentials $V_+$ and $V_-$ for
the case in which an instability is present. An example of data
that yields this kind of potentials is
$(m=14\,,n=10\,,a_1=32\,,c_1=5\,,c_p=5)$. The unstable modes are
those whose pattern speed $\Sigma_{\psi}$ is negative and approach the
minimum of $V_+$ from above. Thus, they are nearly bound states of
the potential well in $V_+$ that can however tunnel out to
infinity through $V_-$. Choosing $\lambda=0$, the potentials $V_+$
and $V_-$ approach zero as $x \rightarrow \infty$, which makes a
tunnelling through $V_-$ easier. } \label{fig:potential}
\end{figure}

Now it is useful to introduce the polynomial
\begin{eqnarray}
 {\cal P}=Bx^3+(A+B)x^2+(\varrho ^2+A)x\,,
\end{eqnarray}
which is positive definite in the range of interest (positive
$x$). We also define
\begin{eqnarray}
 T=-\frac{U_{\rm eff}}{m_{\psi}^{\,2}} \,, \quad
A \equiv \frac{r_+ ^2+M(s_1^2+s_5^2+c_p^2)}{R^2}\,,\quad B \equiv
\frac{r_+ ^2-r_-^2}{R^2}\,.
\end{eqnarray}
Then, we can write the effective Schr$\ddot{\rm o}$dinger equation
(\ref{Schrod 1}) as
\begin{eqnarray}
 \partial ^2 _x H + m_{\psi}^{\,2}\, T\,H=0 \label{Schrod 2}\,,
\end{eqnarray}
with
\begin{equation}
 T=\frac{\cal P}{4x^2(1+x)^2} {\biggl [}\Sigma_{\psi} ^2+\frac{2\varrho m x}{{\cal P}}\Sigma_{\psi}
-\frac{x^2-x(m^2-n^2-1)+n^2}{\cal P} {\biggr ]}\,,
 \label{Def T}
\end{equation}
where we have dropped certain small contributions to $T$.\footnote{More precisely,
we have dropped a term
$1/(m_{\psi}^{\,2} {\cal P})$. This remains a very good
approximation in the high-$m_{\psi}$ limit in which we are
working. As an example, for $n=10$ and $m_{\psi}=10$ the factor
that we dropped is $10^{-4}$ smaller than the last term of
(\ref{Def T}).} Now it is straightforward to factorize the
potential $T$ and write it in the form
\be T=\frac{\cal P}{4x^2(1+x)^2}(\Sigma_{\psi}-V_+)(\Sigma_{\psi}-V_-)
\label{factorize T}\,,\ee
with
\begin{eqnarray}
 V_{\pm}=-\frac{\varrho m x}{{\cal P}} \pm
 {\biggl [} \left ( \frac{\varrho m x}{{\cal P}} \right )^2+
 \frac{x^2-x(m^2-n^2-1)+n^2}{\cal P} {\biggr ]}^{\frac12} \label{potentials}.
 \end{eqnarray}
For general $m\,,n$ the behavior of the potentials $V_+$ and $V_-$
 (see Fig. \ref{fig:potential})  is exactly the same as the one studied
in \cite{cominsschutz}, so we do expect an instability to arise,
as will be shown below. However, and this is a key point, for the
case $m=n+1$ which is the supersymmetric case, we have
\begin{eqnarray}
 V_{+}=-\frac{\varrho m x}{{\cal P}} +
 {\biggl [} \left ( \frac{\varrho m x}{{\cal P}} \right )^2+
 \frac{(x-n)^2}{\cal P} {\biggr ]}^{\frac12}\,,\qquad m=n+1\,,
 \label{potential SUSY}
\end{eqnarray}
which is always positive. Thus, this WKB analysis indicates that
the supersymmetric solutions are stable, as expected.

Hence our radial equation has been reduced to the Schr$\ddot{\rm
o}$dinger form (\ref{Schrod 2}) with an interesting potential
(\ref{factorize T}), which depends on the pattern speed
(\ref{Sigma}). Now the problem becomes to tune this potential by
adjusting $\Sigma_{\psi}$ in order that a `zero-energy' solution can be
found with the appropriate boundary conditions: regular at the
origin and outgoing waves at infinity. Note that in a region
where $\Sigma_{\psi}$ is above $V_+$ or below $V_-$ (allowed regions),
the solutions have an oscillatory behavior. In those intervals
where $\Sigma_{\psi}$ is in between the curves of  $V_+$ and $V_-$
(forbidden regions), the solutions have a real exponential
behavior.

We proceed following \cite{cominsschutz} and study the scattering
of waves in the effective potential constructed above. Consider a
wave that comes from infinity with an amplitude $C_{\rm in}$,
scatters in the ergoregion and returns to infinity with an
amplitude $C_{\rm out}$. In particular, we introduce the
scattering amplitude defined as
\begin{eqnarray}
S\equiv \frac{C_{\rm out}}{C_{\rm in}}.
 \label{Def S}
\end{eqnarray}
The presence of a pole in $S$ (\ie of a resonance) signals the
existence of an instability. Indeed, a pole in $S$ occurs when
$C_{\rm in}=0$ and $C_{\rm out} \neq 0$, and this means that we
have finite outgoing radiation for zero incoming radiation. Near
the pole frequency $\omega_{\rm p}$, the scattering amplitude can
be written to lowest order as \cite{cominsschutz}
\begin{eqnarray}
S\simeq e^{i 2\delta_0}\, \frac{\omega-\omega_{\rm
p}^{\ast}}{\omega-\omega_{\rm p}} \,,
 \label{Def S near pole}
\end{eqnarray}
where $\delta_0$ is a constant scattering phase shift and
$\omega_{\rm p}^{\ast}$ is the complex conjugate of
 $\omega_{\rm p}$. Note that this expression guarantees that
when the frequency of the wave is real, one has
$S(\omega)[S(\omega)]^{\ast}=1$ as required by energy
conservation. Generically, we can write the pole or resonant
frequency as
\begin{eqnarray}
 \omega_{\rm p}=\omega_{\rm r}+i/\tau \,, \label{frequency}
\end{eqnarray}
where $\omega_{\rm r}$ and $1/\tau$ are, respectively, the real
and imaginary parts of $\omega_{\rm p}$. With this convention, a
mode with positive $\tau$ represents an instability, and $\tau<0$
represents a damping mode, since the time dependence\footnote{Our
conventions differ slightly from those of \cite{cominsschutz}.
There waves carry a time dependence $e^{i\omega t}$ while we
follow \cite{ross} which introduces the separation ansatz
(\ref{separation ansatz}) with a time dependence $e^{-i\omega
t}$.} of the resonant wave is given by $e^{-i\omega_{\rm p}
t}=e^{-i\omega_{\rm r} t}e^{t/\tau}$. We can then write
\begin{eqnarray}
S\simeq e^{i 2\delta_0}\,
 \frac{\omega-\omega_{\rm r}+i/\tau} {\omega-\omega_{\rm r}-i/\tau} \,.
  \label{Def S tau}
\end{eqnarray}

To relate the amplitudes $C_{\rm in}$ and $C_{\rm out}$ we apply a
WKB analysis. As we shall learn later on, the unstable modes are
those whose pattern speed $\Sigma_{\psi}$ is negative and approaches the
minimum of $V_+$ from above (see Fig. \ref{fig:potential}). The
scattering problem has then four distinct regions, namely: I, the
innermost forbidden region ($0<x<x_0$); II, the allowed region
where $V_+$ is below $\Sigma_{\psi}$ ($x_0<x<x_1$); III, the potential
barrier region where $V_+$ is above $\Sigma_{\psi}$ ($x_1<x<x_2$); and
finally the external allowed region where $\Sigma_{\psi}$ is below $V_-$
($x_2<x<\infty$). The unstable modes are those that have $\Sigma_{\psi}
<0$. Thus, they are nearly bound states of the potential well in
$V_+$ that can however tunnel out to infinity through $V_-$. In
region I, the WKB wavefunction that vanishes at the origin $x=0$
is
\begin{eqnarray}
H_{\rm I}\simeq \frac{C_1}{m_{\psi}^{1/2}|T|^{1/4}} {\rm
exp}\left[ -m_{\psi}\int_x^{x_0} \sqrt{|T|}\,dx \right]\,,
\label{H1}
\end{eqnarray}
where $C_{1}$ is an amplitude constant. Then, the usual WKB
connection formulae and WKB wavefunctions allow us to relate
$H_{\rm I}$ with the wavefunctions of the other regions and, in
particular, with the incoming and outgoing contributions of the
wavefunction $H_{\rm IV}$ in region IV, which can be written as
\begin{eqnarray}
H_{\rm IV}\simeq \frac{C_6}{m_{\psi}^{1/2}T^{1/4}} {\rm exp}\left[
i\,m_{\psi}\int_{x_2}^{x} \sqrt{T}\,dx \right]
+\frac{C_7}{m_{\psi}^{1/2}T^{1/4}} {\rm exp}\left[
-i\,m_{\psi}\int_{x_2}^{x} \sqrt{T}\,dx \right]\,. \label{H4}
\end{eqnarray}
The WKB analysis yields the relation between the amplitudes $C_6$,
$C_7$ and $C_1$ (see Appendix \ref{sec:A0}):
\begin{eqnarray}
C_1 e^{i \gamma}&=& \frac{1}{2} \left [ \left ( 2\eta
+\frac{1}{2\eta}\right )C_6 +i \left ( 2\eta
-\frac{1}{2\eta}\right )C_7 \right ] \nonumber \\
C_1 e^{-i \gamma}&=& \frac{1}{2} \left [ -i \left ( 2\eta
-\frac{1}{2\eta}\right )C_6 + \left ( 2\eta +\frac{1}{2\eta}\right
)C_7 \right ]\,,
 \label{connectionWKB}
\end{eqnarray}
where
\begin{eqnarray}
\gamma&\equiv& m_{\psi}\int_{x_0}^{x_1} \sqrt{T}\,dx
-\frac{\pi}{4} \,, \label{WKBparameters1}
\end{eqnarray}
\begin{eqnarray}
\ln \eta &\equiv& m_{\psi} \int_{x_1}^{x_2} \sqrt{|T|}\,dx\,.
 \label{WKBparameters2}
\end{eqnarray}
The identification of the ingoing and outgoing contributions in
(\ref{H4}) depends on the sign of $\Sigma_{\psi}$. Indeed, one has
$\Psi\propto e^{-i\omega t} H_{\rm IV}(x)$. If $\Sigma_{\psi}$ is
negative the term $C_6 e^{-i(\omega t-\gamma(x))}$ represents the
ingoing contribution, while the term $C_7 e^{-i(\omega
t+\gamma(x))}$ describes the outgoing contribution (if $\Sigma_{\psi}>0$,
the terms proportional to $C_6$ and $C_7$ in $H_{\rm IV}(x)$
represent, respectively, the outgoing and ingoing modes).
Henceforth we consider the $\Sigma_{\psi}<0$ case (since this will be the
unstable case), for which the scattering amplitude can be written
as
\begin{eqnarray}
S=\frac{C_7}{C_6}=\frac{i(4\eta^2-1)e^{i \gamma}+(4\eta^2+1)e^{-i
\gamma}}{(4\eta^2+1)e^{i \gamma}-i(4\eta^2-1)e^{-i \gamma}}\,.
 \label{WKB S}
\end{eqnarray}
The resonance peaks in the scattering amplitude  occur at a
frequency $\omega_N$ for which $e^{-i \gamma}+ie^{i \gamma}=0$,
\ie when $\gamma(\omega)=\gamma_N$ where
\begin{eqnarray}
\gamma_N (\omega_N) \equiv N\pi +\frac{\pi}{4}\,
 \label{ressonance freq}
\end{eqnarray}
with $N$ being an integer usually referred to as the `harmonic'.
The easiest way to see that the resonance peaks must be near these
(real) frequencies is to note that $S(\gamma_N)=-i$ while for
$\eta\rightarrow \infty$, one has  $S(\gamma \neq \gamma_N)=+i$.
So when $\eta\rightarrow \infty$, one has generally
$S(\gamma)=+i$, but when $\gamma=\gamma_N$ a peak occurs that
changes the value of $S$ from $+i$ to $-i$.

We can now do a Taylor expansion of the functions that appear in
$S$ around $\gamma=\gamma_N$. Defining
\begin{eqnarray}
\alpha = \frac{d \gamma}{d\omega}{\biggl |}_{\omega=\omega_N}=
\frac{d}{d\Sigma_{\psi}} \left [\int_{x_0}^{x_1} \sqrt{T}\,dx
\right]_{\Sigma_{\psi}=\Sigma_{\psi, N} }\,,
 \label{WKB taylor alpha}
\end{eqnarray}
the scattering amplitude can be written as
\begin{eqnarray}
S \simeq \frac{-\alpha(\omega-
\omega_N)+\frac{1}{4\eta^2}-i\left[\alpha(\omega-
\omega_N)+\frac{1}{4\eta^2}\right]}{-\alpha(\omega-
\omega_N)+\frac{1}{4\eta^2}+i\left[\alpha(\omega-
\omega_N)+\frac{1}{4\eta^2}\right]}
 \label{WKB S2}
\end{eqnarray}
which, using $(1+i)/(1-i)=i$, can be cast in the form
\begin{eqnarray}
& & S \simeq i\,\frac{\omega- \omega_N +i \frac{1}{4\eta^2
\alpha}}{\omega- \omega_N -i \frac{1}{4\eta^2 \alpha}}\,.
 \label{WKB S4}
\end{eqnarray}
This result takes the form (\ref{Def S tau}). Hence the discrete
spectrum of resonance frequencies $\omega_N$ is selected by
condition (\ref{ressonance freq}). Further comparing (\ref{Def S
tau}) with (\ref{WKB S4}), one has that the growth or damping
timescale is given by
\begin{eqnarray}
\tau= 4\eta^2 \alpha\,.
 \label{tau}
\end{eqnarray}
Now, $\alpha$ defined in (\ref{WKB taylor alpha}) is always
positive since as $\Sigma_{\psi}$  increases so does $T$ and $\gamma$
defined in (\ref{WKBparameters1}) (the area of the region in
between the $\Sigma_{\psi}$ line and the $V_+$ curve, and in between
$\Sigma_{\psi}$ line and the $V_-$ curve both increase when $\Sigma_{\psi}$
increases). So, we are guaranteed to have a positive $\tau$ and
thus the negative $\Sigma_{\psi}$ modes are unstable. If we redo the
computations to consider the $\Sigma_{\psi}>0$ case, the only difference
is that in (\ref{H4}) the ingoing and outgoing waves are given
instead by the terms proportional to $C_7$ and $C_6$,
respectively. This changes the scattering amplitude from $S$ to
$S^{-1}$ and thus $\tau$ to $-\tau$ implying that the positive
$\Sigma_{\psi}$ modes are damped.

Though the resonance frequencies and growth timescales can be computed
with numerical methods from (\ref{ressonance freq}) and
(\ref{tau}), as we shall do in Section \ref{num res}, we can
still make some further progress analytically by approximating
the well of $V_+$ by a parabola. Near the well, the potential
$V_+$ behaves generally as
\begin{eqnarray}
 V_+ \simeq \frac{(x-x_{\rm m})^2}{P_{\rm m}}+a_{\rm m}\,,
\label{parabola}
\end{eqnarray}
with $a_{\rm m}<0$. The boundaries $x_0$ and $x_1$ are the roots
of $\Sigma_{\psi}-V_+$, namely: $x_0=x_{\rm m}-[P_{\rm m}(\Sigma_{\psi}-a_{\rm
m})]^{1/2}$ and $x_1=x_{\rm m}+[P_{\rm m}(\Sigma_{\psi}-a_{\rm
m})]^{1/2}$. Since $\sqrt{T}$ vanishes at these boundaries one has
\begin{eqnarray}
\alpha = \int_{x_0}^{x_1} \frac{d \sqrt{T} }{d\Sigma_{\psi}}\,dx \,.
 \label{alpha 2}
\end{eqnarray}
Moreover, near the bottom of the well, only $\Sigma_{\psi}-V_+$ varies
significantly with $x$, and we can assume that all the other
quantities that appear in the integral of $\alpha$ are
approximately constants given by their value at $x=x_{\rm m}$ (the
accuracy of this assumption increases as $\Sigma_{\psi}$ approaches
$a_{\rm m}$). One then has
\begin{eqnarray}
\alpha \simeq \frac{\Sigma_{\psi}+\frac{\varrho m x_{\rm m}}{{\cal P}(x_{\rm
m})} } { \sqrt{\Sigma_{\psi} - V_-(x_{\rm m})} } \frac{\sqrt{{\cal
P}(x_{\rm m})}} {2x_{\rm m}(1+x_{\rm m})} \int_{x_0}^{x_1} \left
[\Sigma_{\psi}-V_+ \right ]^{-\frac12}\,dx \,,
 \label{alpha 3}
\end{eqnarray}
with $V_+$ given by (\ref{parabola}), which yields for $\alpha$
the value
\begin{eqnarray}
 \alpha =\pi \sqrt{P_{\rm m}}\left [\Sigma_{\psi}+\frac{\varrho m x_{\rm
m}}{{\cal P}(x_{\rm m})} \right ]\left [\Sigma_{\psi} - V_-(x_{\rm m})
\right ]^{-1/2}\frac{\sqrt{{\cal P}(x_{\rm m})}} {2x_{\rm
m}(1+x_{\rm m})} \label{alpha 4}\,.
\end{eqnarray}

Let us illustrate the use of the WKB method we have described in
this section to compute the instability parameters in a particular
configuration. Take,
\begin{eqnarray}
m&=&14\,;\,\,n=10\,;\,\,a_1=32\,;\,\,c_1=5\,;\,\,c_p=5\,; \label{house} \\
& & \lambda=m_{\phi}=0\,;\,\,l=m_{\psi}=10\,.\label{house2}
 \end{eqnarray}
By approximating the well in $V_+$ by a parabola, as in
(\ref{parabola}), we get
\begin{eqnarray}
 a_{\rm m}=-0.17894\,;\,\,x_{\rm m}=9.1537\,;\,\,P_{\rm
m}=2759.4\,.
\end{eqnarray}
The resonant frequencies are those that satisfy condition
(\ref{ressonance freq}) with $\gamma(\omega)$ given by
(\ref{WKBparameters1}). For the fundamental harmonic ($N=0$), we
get
\begin{eqnarray}
\Sigma_{\psi}=-0.173 \,.
 \label{numeric sigma 0}
\end{eqnarray}
The growth timescale of the instability is given by (\ref{tau})
with $\eta(\omega_N)$ given by (\ref{WKBparameters2}). Again, for
$N=0$ we get
\begin{eqnarray}
 \tau \sim 10^{47}\,.
 \label{numeric tau}
\end{eqnarray}

Independently of the parameters of the geometry, we note that as
$m_{\psi}$ grows, $\Sigma_{\psi}$ approaches $a_{\rm m}$, the value of
the $V_+$ at its minimum. For the particular geometry parameters
described in  (\ref{house}) we have (for $\lambda=m_{\phi}=0$):
\begin{eqnarray}
 m_{\psi}&=&10:\,\,\Sigma_{\psi}=-0.173\,, \nonumber \\
m_{\psi}&=&20:\,\,\Sigma_{\psi}=-0.176\,, \nonumber \\
m_{\psi}&=&40:\,\,\Sigma_{\psi}=-0.177\,.
 \label{numeric sigma}
\end{eqnarray}
This feature can be proved analytically, as was  done in
\cite{cominsschutz}.

Let us verify consistency of our results. We have assumed that
$a_i^2\frac{\omega^2-\lambda ^2}{R^2}\ll 1$ in order to do the
approximation $\Lambda \simeq l(l+2)$. Now, for the cases we
listed above one has $ a_i^2\frac{\omega^2-\lambda ^2}{R^2} \sim
10^{-2}$, which is inside the range of validity for our
approximations. A different combination of parameters yields a
different instability timescale, and resonant frequency, so there
are geometries more unstable than others. The following refers to
the fundamental harmonic, and are computed within the parabolic
approximation. We work with the following parameters:
\begin{eqnarray}
c_1=5\,;\,\,c_p=5\,;\,\,\lambda=m_{\phi}=0\,;\,\,
l=m_{\psi}=10\,.\end{eqnarray}
For the fundamental harmonic we then get
\beq m&=&1400\,;\,\,n=10\,;\,\,a_1=32\,;\,\,
\Sigma_{\psi}=-0.40502\,,\,\,\tau \sim 3\times 10^{82} \\
 m&=&12\,;\,\,n=10\,;\,\,a_1=32\,;\,\, \Sigma_{\psi}=-0.104\,,\,\,\tau \sim
3\times 10^{52} \\
m&=&14\,;\,\,n=10\,;\,\,a_1=3200\,;\,\, \Sigma_{\psi}=-0.1728\,,\,\,\tau
\sim 3\times 10^{48} \\ m&=&3\,;\,\,n=1\,;\,\,a_1=32\,;\,\,
\Sigma_{\psi}=-0.0148\,,\,\,\tau \sim 3\times 1.7\times10^{44} \eeq
It also evident that the instability is much stronger for small
values of $m_{\psi}$, where the WKB is expected to break down.

To conclude this section, let us consider the regime of validity
of the WKB approximation with more detail. A standard analysis of
Eq.~(\ref{Schrod 1}) suggests the WKB approximation is valid for
$\left |\partial_xU_{\rm eff} \right | \ll |U_{\rm eff}|^2$, which
can be rewritten as $\left |
\partial_x T /T^2(x) \right | \ll m_{\psi}^2$. So, for large
$m_{\psi}$, the WKB approximation seems to be valid quite
generally. However, we must sound a note of caution. As we already
remarked, Eq.~(\ref{numeric sigma}) shows that as $m_{\psi}$
grows, $\Sigma_{\psi}$ approaches $a_{\rm m}$, the value of the $V_+$ at
its minimum
--- this can be proved analytically \cite{cominsschutz}. So when
$m_{\psi}$ becomes very large, the two turning points are very
close and the WKB analysis breaks down because $T(x)\rightarrow
0$. So we conclude that the WKB approximation used in this section
should be valid in a regime with large $m_{\psi}$, but not exceedingly
large. In any event, it is clear that the instability is strongest
for small values of $m_{\psi}$, when the WKB analysis is certainly
not valid. So, in the next two sections we will compute the
features of the instability using complementary methods valid for
small values of $m_{\psi}$. We will also find remarkable agreement
between all three approaches.


\section{\label{sec:Match}Matched
asymptotic expansion analysis}

The WKB analysis described in the last section appears to be
strongest when describing solutions for which $\kappa ^{-1}\sim
\zeta,\xi$, but in general this corresponds to solutions with high
angular momentum.  In the sense that the timescale of the
instability due to these modes is largest, they are the least
unstable. Conversely, the matched asymptotic expansion that we use
in this section becomes valid when $\kappa ^{-1}> \zeta,\xi$, they
are the dominant decay modes. As an additional bonus, the
eigenvalues are determined explicitly through algebraic
constraints. Having both approximations at our disposal allows us
to accurately calculate the eigenvalues for most of the allowed
parameters.

We follow a matching procedure introduced in \cite{staro1}, which
has previously been used for studying scalar fields in three-charge
geometries by Giusto, Mathur and Saxena \cite{three2}, in the
\jmart backgrounds \cite{ross} and also in \cite{bhb,detweiler,bhbAdS,cardoso,CardDiasYosh}.
The space is divided into two parts: a near-region, $x\ll \beta$, and a
far-region, $x\gg \alpha$, such that $\alpha \ll \beta$.  The
radial equation is then solved approximately and the appropriate
boundary conditions applied in each of the two regions.  Finally,
we match the near-region and the far-region solutions in the area
for which they are both valid, $\alpha \ll r \ll \beta$. This
gives a set of constraints, the solution of which gives the
eigenvalues.  Performing this analysis for the radial equation
(\ref{rad eq0}), we shall see that the only solutions which are
regular at the origin and purely outgoing at infinity are finite
as $x \rightarrow \infty$, and lead to instabilities. Except when
otherwise stated, the analysis in this section will hold for
general values of $m_{\psi}$, $m_{\phi}$ and $\lambda$.

\subsection{\label{sec:BH Near region}The near region solution}

In the near-region, $\kappa^{2} x\ll |1-\nu^2|$, one can neglect the
$\kappa^{2}x$ term, and the radial equation (\ref{rad eq0}) is
approximated by
\begin{equation}
 x(1+x)\partial_x^2 h+ (1+2x)
\partial_x h +\frac{1}{4} \left [ 1-\nu^2+
\frac{\xi^2}{x+1}-\frac{\zeta^2}{x}\right ] h=0\,.
 \label{near wave eq}
\end{equation}
With the definition $h=x^{|\zeta|/2} (1+x)^{\xi/2}\,w$, the
near-region radial equation becomes a standard hypergeometric
equation \cite{abramowitz} of the form
\begin{equation}
x(1+x)\partial_x^2 w+[c+(a+b+1)x]\partial_x w+ab \, w=0,
\end{equation}
where
\begin{equation}
a=\frac{1}{2}\left (1+|\zeta|+\xi+\nu \right ) \,, \qquad
b=\frac{1}{2}\left (1+|\zeta|+\xi-\nu \right )\,, \qquad c=1+
|\zeta| \,.
 \label{hypergeometric parameters}
\end{equation}
The full solution to the above is given in terms of hypergeometric
functions as $w = A\, F(a,b,c,-x)+B\, x^{1-c}
F(a-c+1,b-c+1,2-c,-x)$, which allows us finally to write the
solution of the radial equation in the near region as
\begin{eqnarray}
h &=& A\,x^{|\zeta|/2}(1+x)^{\xi/2} F(a,b,c,-x) \nonumber \\
& & +B \, x^{-|\zeta|/2}(1+x)^{\xi/2} F(a-c+1,b-c+1,2-c,-x) \,.
\label{hypergeometric solution}
\end{eqnarray}
At this point we impose the first boundary condition: the solution
must be regular at $x=0$ since the geometry is smooth at the
origin of the ``core". The term proportional to $x^{-|\zeta|/2}$
diverges at $x=0$ and must be discarded, \ie its coefficient, $B$,
must be set to zero.

To perform the matching we need to know the large $x$ behavior
behavior of the regular near-region solution. To this end, one uses
the $x \rightarrow 1/x$ transformation law for the hypergeometric
function \cite{abramowitz}
\begin{eqnarray}
F(a,b,c,-x)&=&\frac{\Gamma(c)\Gamma(b-a)}{\Gamma(b)\Gamma(c-a)}\,
x^{-a}
 \,F(a,1\!-\!c\!+\!a,1\!-\!b\!+\!a,-\!1/x)  \nonumber \\
& & + \frac{\Gamma(c)\Gamma(a-b)}{\Gamma(a)\Gamma(c-b)}\,x^{-b}
 \,F(b,1\!-\!c\!+\!b,1\!-\!a\!+\!b,-\!1/x)\,,
 \label{transformation law}
\end{eqnarray}
and the property $F(a,b,c,0)=1$. Note that this expression for
the transformation is only valid when $a-b=\nu$ is non-integer.  This is an assumption
we will continue to make throughout this section.  In the end, we
shall derive a condition determining the allowed eigenvalues that
will not be dependent upon this assumption and therefore we may
extend our results to integer values of $\nu$ by continuity.

The large $x$ behavior of the near-region solution is then given
by
\begin{eqnarray}
& & \hspace{-0.5cm} h \sim A\,\Gamma(1+|\zeta|)  {\biggl [}
\frac{\Gamma(-\nu)}
 { \Gamma\left [\frac{1}{2}\left (1+|\zeta|+\xi-\nu \right )
 \right ]
 \Gamma\left [\frac{1}{2}\left (1+|\zeta|-\xi-\nu \right )
 \right ] }\:
x^{-\frac{\nu+1}{2}}\nonumber \\
& & \hspace{2 cm}
 +\frac{\Gamma(\nu)}
{ \Gamma\left [\frac{1}{2}\left (1+|\zeta|+\xi+\nu \right )
 \right ]
 \Gamma\left [\frac{1}{2}\left (1+|\zeta|-\xi+\nu \right )
 \right ] }\:
x^{\frac{\nu-1}{2}} {\biggr ]}.\nonumber \\
& &
 \label{near field large r}
\end{eqnarray}

\subsection{\label{sec:BH Far region}The far region solution}

In the far-region, $\kappa x^2\gg {\rm max}\{ \xi^2-1,\zeta^2 \}$, the
terms $\xi^2/(x+1)$ and $\zeta^2/x$ can be neglected, and the
radial equation can be approximated by
\begin{eqnarray}
\partial_x^2 (x h)+ \left [ \frac{\kappa^{2}}{4x}-\frac{\nu^2-1}{4x^2}
\right ] (x h)=0\,.
 \label{far wave eq}
\end{eqnarray}
The most general solution of this equation when $\nu$ is
non-integer is a linear combination of Bessel functions of the
first kind \cite{abramowitz},
\begin{eqnarray}
h=x^{-1/2}\left [ C J_{\,\nu}(\kappa \sqrt{x})+ D
J_{\,-\nu}(\kappa \sqrt{x})\right ]\,.
 \label{far field}
\end{eqnarray}
This form does not lend itself easily to application of the
boundary conditions. Instead, for large $\kappa \sqrt{x}$, the
solution may be expanded as \cite{abramowitz}
\begin{eqnarray}
 h \sim \frac{x^{-3/4}}{\sqrt{2\pi \kappa}}
{\biggl [} e^{i\kappa \sqrt{x}} e^{-i\frac{\pi}{4}}\left (C e^{-i
\frac{\pi\nu}{2}} +D e^{i \frac{\pi\nu}{2}} \right )
 + e^{-i\kappa \sqrt{x}}
e^{i \frac{\pi}{4}}\left (C e^{i \frac{\pi\nu}{2}} +D e^{-i
\frac{\pi\nu}{2}} \right ){\biggr ]}\,.
 \label{far field-large r}
\end{eqnarray}
As in the WKB analysis, we assume that the real part of $\omega$
is negative, and therefore the positive and negative sign
exponentials give, respectively, ingoing and outgoing waves.  We
require that there be purely outgoing waves at infinity and so
impose the constraint that the coefficient of the positive
exponential vanishes, yielding
\begin{eqnarray}
C = -D e^{i \pi\nu}. \label{far amplitude relation}
\end{eqnarray}
When $\omega$ becomes complex, so too does $\kappa$. Since the
sign of the real part of $\omega$ is negative, the definition of
$\kappa$ (\ref{rad eq parameters0}) implies that its imaginary
part has a sign opposite that of the imaginary part of $\omega$.
Therefore, requiring additionally that the solution be finite as
$x\rightarrow \infty$ implies that the imaginary part of $\omega$
must be positive.  This is precisely the sign for the imaginary
part of the frequency that leads to instabilities.  Thus we see
that simply requiring the solutions with complex frequency be
finite at infinity automatically guarantees they lead to
instabilities.

Now, to do the matching in the overlapping region, we will need to
know how the far-region solution behaves for small values of $x$.
More specifically, for small $\kappa \sqrt{x}$, and considering
only the dominant terms, the solution behaves as \cite{abramowitz}
\begin{eqnarray}
h \sim D\left [ \frac{(2/\kappa)^{-\nu}}{\Gamma(1+\nu)}\:
x^{\frac{\nu-1}{2}} - e^{i \pi\nu}
\frac{(2/\kappa)^{\nu}}{\Gamma(1-\nu)}\: x^{-\frac{\nu+1}{2}}
\right ].
 \label{far field-small r}
\end{eqnarray}

\subsection{\label{sec:MatchCond}Matching conditions: Selection of
frequencies}

We will now determine the frequencies that can appear when the
geometry is perturbed by a scalar field.  The frequency spectrum
is not arbitrary: only those values that satisfy the matching
conditions between the near-region and the far-region are allowed.
We shall see that there are two solutions of the matching
equations, yet only one will lead to instabilities.

Matching the powers of $x$ between the near (\ref{near field large
r}) and far-region solutions (\ref{far field-small r}), and taking
a ratio to eliminate the amplitudes $A$ and $D$, yields
\begin{equation}
 -e^{i \pi \nu} (\kappa/2)^{2\nu}
\frac{\Gamma(1-\nu)}{\Gamma(1+\nu)}
=\frac{\Gamma(\nu)}{\Gamma(-\nu)}
\frac{\Gamma(\frac{1}{2}(1-\nu+|\zeta|+\xi))}{\Gamma(\frac{1}{2}(1+\nu+|\zeta|+\xi))}
\frac{\Gamma(\frac{1}{2}(1-\nu+|\zeta| -
\xi))}{\Gamma(\frac{1}{2}(1+\nu+|\zeta|-\xi))} \ .
\label{eq:asymp2_finaleq}
\end{equation}
The problem of finding the outgoing modes thus boils down to
solving the single transcendental equation
(\ref{eq:asymp2_finaleq}); we will do so by iteration.  Note that
the $\kappa$ dependence on the left hand side means that it is
suppressed.  For the equation to hold, a similar suppression must
also occur on the right hand side.  This is only possible if one
of the gamma functions in the denominator of the right side is
large. Since the gamma function diverges when its argument is a
non-positive integer, we take as a first iteration the choice
\begin{equation}
\nu+|\zeta| - \xi = -(2 N + 1) \ , \label{eq:asymp2_quantization}
\end{equation}
where  the non-negative integer $N$ will again be referred to as
the harmonic.  Note that we could also have chosen the above
relation, but with the opposite sign for $\xi$.  While this does indeed
lead to a solution, one finds that the imaginary part of the frequency
is always negative, \ie the modes are exponentially damped in time.

This first estimate is obviously not the end of the story as it
would cause the right side to completely vanish. To go beyond this
approximation, we rewrite Eq.~(\ref{eq:asymp2_finaleq}) in terms
of $N$, then perturb $N \rightarrow N + \delta N$, where $\delta N
\ll N$. This deformation appears at leading order only for the
$\Gamma$ function in the denominator on the right hand side that
diverges, it may be neglected in all other factors. More
concretely, to extract $\delta N$ from the $\Gamma$ function we
use $\Gamma(z)\Gamma(1-z)= \pi/\sin(\pi z)$, and sine function
identities to obtain the expansion
\begin{equation}
\Gamma(-N-\delta N) \approx  -\left [ (-1)^N N! \thinspace \delta
N \right ]^{-1}.
\end{equation}
Substituting this into (\ref{eq:asymp2_finaleq}), and using
a number of $\Gamma$ function identities, we solve
for the imaginary part of the first correction
\begin{equation}
{\mathcal I m}(\delta N) = \pi \frac{ (\kappa/2)^{2\nu} }{\Gamma^2(\nu)}
    [\nu]_N \thinspace [\nu]_{N+|\zeta|} \ , \label{deltaN}
\end{equation}
where $[a]_n = \prod_{i=1}^n (1+a/i)$.  Since $N$ is ${\mathcal O}(1)$
and $\delta N \sim \kappa^{2 \nu}$, we see that we may stop after the first
iteration.  As a function of $\nu$, this can have a single
maximum near $\nu \sim \kappa$.  In general we will have $\kappa \ll 1$ and
$\nu \sim 1+l$, so will always be in a region where this is
a monotonically decreasing function of $\nu$.  For fixed $\nu$,
the last two factors make this an increasing function of $N$ and
$|\zeta|$, but the general behavior will be dominated by the
effects of changing $\nu$.

The equation (\ref{eq:asymp2_quantization}) uniquely determining
$\omega$ can be exactly solved
\begin{equation}
(\varepsilon + \varrho^2) \thinspace \omega = - \left(\lambda
\frac{s_p c_p M}{R^2}+\varrho c \right ) + \sqrt{ \left(\lambda
\frac{s_p c_p M}{R^2}+\varrho c \right
)^2-(\varepsilon+\varrho^2)(c^2-\nu_0^2) } \ , \label{omega_solution}
\end{equation}
where
\begin{equation}
\varepsilon  \equiv  \frac{1}{R^2}(r_+^2 + M(s_1^2 + s_5^2 +
c_p^2)) \ , \qquad c   \equiv   \xi_0 - |\zeta| - (2N+1) \ ,
\end{equation}
and a variable with a subscripted $0$ means we have set
$\omega=0$.  Note that as long as $m \geq n+2$, we have
$\varepsilon/\varrho^2 \ll 1$ and both quantities are positive.  When
$m \rightarrow n+1$, though, $\varepsilon \rightarrow -\infty$
(since $M\rightarrow 0$, $r_+^2 \rightarrow -\infty$ and $R^2$
stays finite), ensuring that there can be no instability for the
supersymmetric solutions. This extends to arbitrary modes the
conclusion from the discussion associated to equation (\ref{potential SUSY})
for modes with $m_\phi=\lambda=0$.

When evaluated on a solution, $\nu$ is given by $\nu=\omega \varrho +
c$. Since we are interested in solutions for which $\omega$ is
negative, this means $c>0$.  Then, requiring that $\omega$ be
negative and real, gives three more conditions.  The first ensures
that the result is real while the second requires that the first
term of (\ref{omega_solution}) is negative.  Finally, the
condition that appears to be the most difficult to satisfy ensures
the contribution from the square root does not make the total
result positive, \ie
\begin{equation}
c^2-\nu_0^2 > 0 \ . \label{constraint}
\end{equation}
When $\lambda \neq 0$, these conditions must also be supplemented by the requirement
that $\omega^2-\lambda^2>0$, which ensures the asymptotic behavior of the solution is correct.
With these satisfied, we may determine the effect of the correction.

The imaginary contribution to $N$ is taken as resulting from a
small imaginary correction to $\omega$. Then, the two are related
through
\begin{eqnarray}
\delta N & = & \left . \frac{\delta \omega}{2} \frac{d}{d\omega} (\xi-\omega) \right |_N \nonumber \\
& = & \frac{\delta \omega}{2\nu}
\left [ (\varrho^2+\varepsilon)\omega+(\lambda s_p c_p M/R^2+\varrho c) \right ] \nonumber \\
& = & \frac{\delta \omega}{2\nu} \sqrt{ \left(\lambda \frac{s_p
c_p M}{R^2}+\varrho c \right )^2-(\varepsilon+\varrho^2)(c^2-\nu_0^2) }
\ .
\end{eqnarray}
In the final line we have used the solution (\ref{omega_solution})
to show that the sign of $\delta N$ determines the sign of the
correction to $\omega$. Since ${\mathcal Im}(\delta N)$ is always
positive when evaluated on the solution of
(\ref{eq:asymp2_quantization}), the corresponding imaginary part
of $\omega$ is positive.

To summarize,  whenever the constraints, in particular
(\ref{constraint}), are satisfied there is a corresponding
outgoing mode of the scalar field equation. Further, the imaginary
part of the frequency of this mode is guaranteed to be positive,
indicating that it leads to an instability.  The timescale for the
instability generated by the mode is a monotonically increasing
function of $\nu$, which is given by
\begin{eqnarray}
(\varepsilon + \varrho^2) \thinspace \nu & = & \varepsilon c -
\lambda \varrho s_p c_p \frac{M}{R^2} +
    \sqrt{ \left ( \varepsilon c - \lambda \varrho s_p c_p \frac{M}{R^2} \right )^2
    + (\varepsilon + \varrho^2)(2 c \varrho \lambda s_p c_p \frac{M}{R^2} + \nu_0^2 \varrho^2 - \varepsilon c^2) }
    \ . \label{nu_solution}
\end{eqnarray}
A similar argument, based on the solution of equation
(\ref{eq:asymp2_quantization}), but with the opposite sign for
$\xi$ would lead to a set of outgoing modes with an amplitude that
decays in time.

As an example, consider the particular background geometry and
scalar field solution described by
\begin{eqnarray}
m&=&5\,;\,\,n=1\,;\,\,a_1=19.1\,;\,\,c_1=5\,;\,\,c_p=1.05\,;\nonumber
\\& &  \lambda=m_{\phi}=0\,;\,\,l=m_{\psi}=2\,.
\end{eqnarray}
 The first two iterations with $N=0$ gives
\begin{eqnarray}
\omega & = & -2.8717 \ , \nonumber \\
\tau^{-1}={\mathcal Im (\delta \omega)} & = & 4.42 \times 10^{-11}
\ ,
\end{eqnarray}

The results obtained here are consistent with the WKB analysis of
the last section, \ie there are outgoing modes that rotate in the
same sense as the background geometry whose amplitude grows
exponentially in time.  What we have gained is an explicit set of
relations that allows the unstable mode frequencies to be
calculated.  In particular, one can now make definite statements
about the relative timescales for unstable modes just by looking at
equation (\ref{nu_solution}). We leave the precise details of this
to Appendix \ref{beyond} and just give the results here. The most
unstable modes are those which minimize $\nu$.  Since $\varepsilon
\ll \varrho^2$ this generally means that the modes which maximize $c$
or minimize $\nu_0$ will be the most unstable.  In general this
means we should consider the lowest possible $l$ for which the
constraints can be satisfied when setting $m_\psi = l$, $m_\phi=0$
and $N=0$.

A second benefit of this analysis is an improvement in accuracy for the most
unstable modes.  For comparison, performing the
WKB analysis and not neglecting any terms in the potentials or approximating the
bottom of the well with a parabola gives $\omega = -3.129+4.00
\times 10^{-10}i$. From the full numerical solution we have
$\omega=2.8718+4.46 \times 10^{-11}i$.  For values of $\omega$ in
this range we have $\kappa^{-2} \sim 1900$, so we are well within
the range for which we should trust this solution. As $\kappa
^{-1}$ approaches $\max(|\zeta|,\xi)$, this analysis begins to break
down, but it appears that the WKB approach becomes increasingly
accurate. In the next section we will present a more detailed list
of eigenvalues corresponding to instabilities and discuss the
results.

\section{\label{numerical} Numerical Results}

We will now solve the radial equation (\ref{rad eq0}) numerically
to extract the instability. We begin with an exposition of the
numerical algorithm. The only approximation used in this section
concerns the angular eigenvalue $\Lambda$, that we shall assume to
be well described by (\ref{app}). At the end of the calculation we
always make sure the result fits the regime of validity of this
approximation.  Note, however, solutions can still be found even
when outside this range.  The easiest
way to do this is by treating $\Lambda$ and $\omega$ respectively as eigenvalues
of the angular and radial equations.  The coupled system may then be solved by
first assuming the approximation to hold and solving the radial equation
for $\omega$, this is then fed into the angular equation to obtain
an improved value of $\Lambda$.  This process may be iterated until
the desired level of convergence is achieved.

\subsection{\label{num proc} Numerical procedure}

The method of finding solutions numerically is very much like
performing the matched expansions.  We use
Eqs.~(\ref{hypergeometric solution}) and (\ref{far field}) to fix
the initial conditions for two integrations of the exact radial
equation.  Since the equation of motion is linear, we may
immediately match the two solutions at a point in the interior
region by rescaling.  This leaves two more conditions to be
satisfied, that matching of the derivatives of the real and
imaginary parts. Fixing all other parameters, we vary the real and
imaginary parts of $\omega$ to satisfy these conditions.

Given the small size of the expected imaginary part, it is most
straightforward to use a package like Mathematica
\cite{mathematica} with its software based arbitrary precision, to
perform the calculations.  Satisfying the matching conditions can
be done by treating the difference in derivatives at the interior
point as a complex valued function of $\omega$.  A root may then
be searched for using the built-in function {\tt FindRoot} which,
for a function without explicit derivatives, looks for the
solution by constructing secants.

Since the imaginary part is expected to be far smaller than the
real, gradients of the matching function in the imaginary $\omega$
direction will be large only when very near a solution, but
negligible elsewhere.  The initial guesses at the solution are
therefore very important for ensuring that iterations converge to
a solution.  It was found empirically that solutions could
consistently be found by choosing to start the search in a region
around the real value of $\omega$ for which the inner solution
vanishes at the matching point.  Small changes in the imaginary
part of $\omega$ near this point appear to be sufficient to bring
about convergence.
\begin{figure}[h]
\begin{center}
\resizebox{9cm}{6cm}{\includegraphics{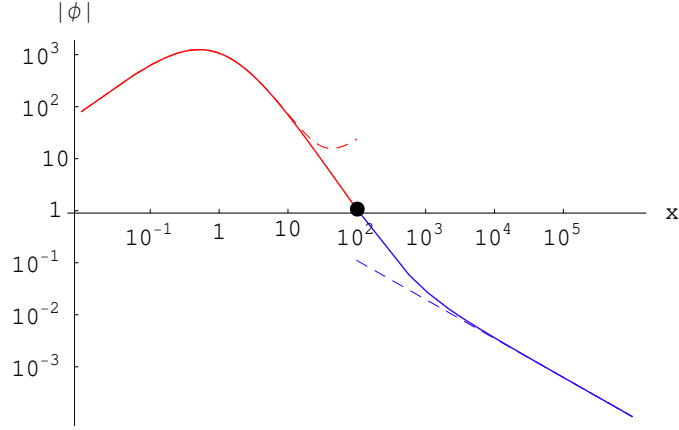}}
\end{center}
\caption{An example solution showing vanishing as both
$x\rightarrow 0$ and $x \rightarrow \infty$.}
\label{fig:sampleplot}
\end{figure}
In Fig.  \ref{fig:sampleplot} we show an example solution obtained
in this manner.  The solid line is the full numeric solution, with
the integration starting at small $x$ in red to the left of the
black dot and that starting at large $x$ on the right in blue. The
dashed lines are the near (\ref{hypergeometric solution}) and far
(\ref{far field}) approximations used to set the initial
conditions for integrating the exact radial equation. The fact
that the imaginary part of $\omega$ is in general very small
raises non-trivial problems, related to the number of digits of
precision used and the exact way in which boundary conditions are
applied. A discussion of these aspects is deferred to Appendix
\ref{sec:A1}.

\subsection{\label{num res} Numerical results}

\begin{figure}[h]
\begin{center}
\resizebox{7cm}{5cm}{\includegraphics{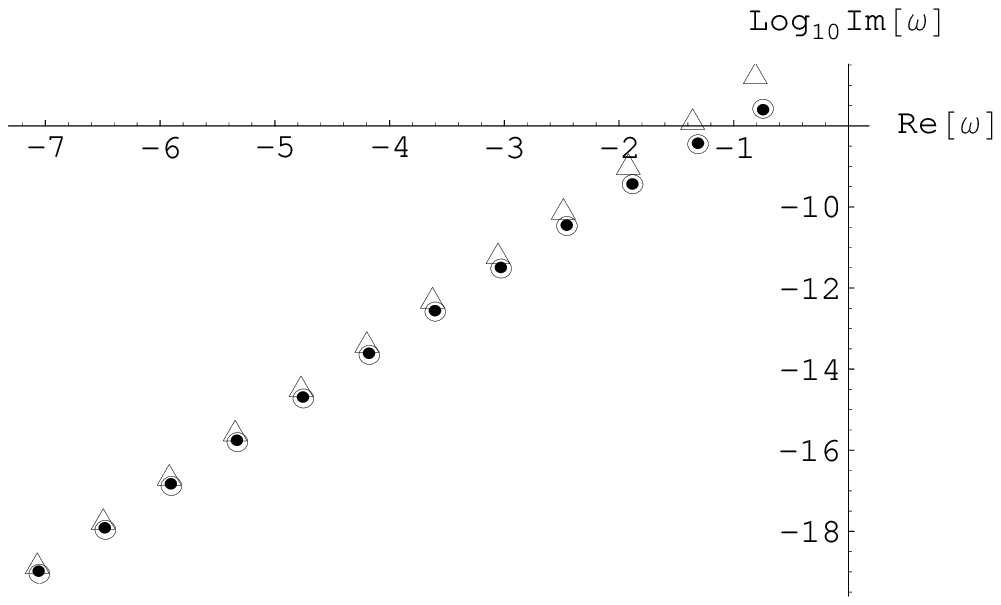}}
\resizebox{7cm}{5cm}{\includegraphics{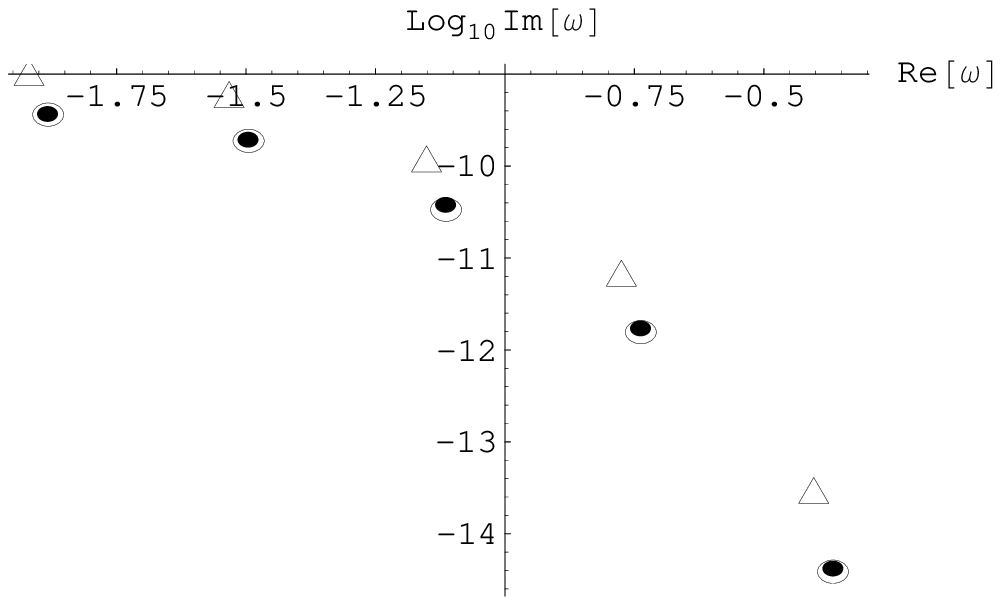}}
\end{center}
\caption{On the left we choose the  lowest harmonic and vary
$l=m_{\psi}$ from 2 to 13 from upper right to lower left.  The
solid circles represent the numeric solutions, while the triangles
are the results of the WKB analysis and the unfilled circles
correspond to the matched expansion.  On the right we fix
$l=m_{\psi}=4$ and vary the harmonic from 0 to 4 from upper left
to lower right.} \label{fig:fixNvaryl_compare_methods}
\end{figure}

Our numerical results are summarized in Fig.
\ref{fig:fixNvaryl_compare_methods} and Table
\ref{tab:efficiency3}. In Fig. \ref{fig:fixNvaryl_compare_methods}
on the left we present the numerical solutions obtained for
\begin{equation}
m = 5\,,\,\, n = 1 \,,\,\, c_1 = 1.1\,,\,\,c_5=1.52\,,\,\, a_1 =
262.7 \,,\,\, \lambda = m_{\phi} = 0 \ .
\end{equation}
where we consider only the lowest harmonic, $N=0$, but vary
$l=m_\psi$. At $l=1$, $ \kappa ^{-1} \sim 40$ indicating the
matched solution is valid, as $l$ a grows so do $\xi, \zeta$ while
$\kappa ^{-1}$ shrinks, meaning the approximation should soon
break down. At $l=5$, $\kappa^ {-2} \sim 10$  and the
approximation is becoming no longer valid. Finally, when $l=13$,
$\kappa^ {-2} \sim 1$ and differences between the matched and
numerically determined eigenvalues are starting to become
apparent.  In Fig. \ref{fig:fixNvaryl_compare_methods} on the
right, we use the same parameters as before, but now fix
$l=m_{\psi}=4$ and vary the harmonic from $N=0$ up to $4$.
Increasing $N$ leads to smaller values of $\omega$ and therefore
smaller values of $\kappa$, so that the matched solutions are
valid throughout. It should also be noted that if the
approximation $a_1^2 \omega^2/R^2 \ll m_{\psi}^2$ is valid for a
given $m_{\psi}$, then it should be valid for all $m_{\psi}$. This
is because $\omega$ scales with $m_{\psi}$, as we observed within
the WKB approximation.

In Table \ref{tab:efficiency3} we present and compare the
numerical results with those obtained through the approximate
analyical approaches. The values labeled as WKB$_{\rm NUM}$
(numerical WKB) stand for values obtained using the full WKB
approximation, formulae (\ref{WKBparameters1}),
(\ref{WKBparameters2}), (\ref{ressonance freq}) and (\ref{tau}),
which have been handled numerically. The values obtained using the
parabolic approximation, formulae
 (\ref{parabola})-(\ref{alpha 4}), for the potential are denoted
 by WKB$_{\rm AN}$.

Notice first that all the different approaches yield consistent
and in fact very similar results: they are all rather accurate in
their own regime of validity. As predicted by the analytic
approaches, and verified numerically, the real part of the
frequency scales with $m_{\psi}$, whereas the logarithm  of the
imaginary part scales with $m_{\psi}$, \eg see
Eq.~(\ref{WKBparameters2}). Thus the instability timescale
increases rapidly as a function of $m_{\psi}$.

\begin{table}
\begin{center}
\begin{tabular}{|c|c|c|c|c|} \hline
$m_\psi$ & \multicolumn{1}{c|}{Numeric} & \multicolumn{1}{c|}{
WKB$_{\rm NUM}$} &\multicolumn{1}{c|}{WKB$_{\rm AN}$}
& \multicolumn{1}{c|}{Matching}  \\
\hline
1  &$ -0.184 +  3.83 \times 10^{-8}i  $& $ - $&$-$ &$ -0.184 + 3.83 \times 10^{-8}i $\\
\hline
 2  &$ -0.744 +  2.51 \times 10^{-8}i  $& $ -0.812 + 1.61 \times 10^{ -7}i $&$-0.826+1.89\times 10^{-7}i$ &$ -0.744 + 2.64 \times 10^{-8}i $\\
\hline
 3  &$ -1.312 +  3.73 \times 10^{ -9}i  $& $ -1.359 + 1.20 \times 10^{ -8}i $& $-1.371+1.48\times 10^{-8}i$&$ -1.312 + 3.53 \times 10^{ -9}i$ \\
\hline
 4  & $-1.883  + 3.69 \times 10^{ -10}i  $&$   -1.919 + 9.33 \times 10^{ -10}i  $&$-1.932+1.17\times 10^{-9}i$ &$ -1.882  + 3.63 \times 10^{ -10}i$ \\
\hline
5   & $-2.456  + 3.55 \times 10^{ -11}i $&$  -2.486 + 7.37 \times 10^{ -11}i  $&$-2.499+9.39 \times 10^{-11}i  $ &$ -2.454 +  3.39 \times 10^{ -11}i $\\
\hline
6  & $ -3.030  + 3.22 \times 10^{ -12}i $& $  -3.055  + 5.89 \times 10^{ -12}i  $& $-3.072+7.62\times 10^{-12}i$ &$ -3.028 +  3.02 \times 10^{ -12}i $\\
\hline
7  & $ -3.605  + 2.77 \times 10^{ -13}i $&$   -3.626   +4.73 \times 10^{ -13}i  $&$-3.647+6.23 \times 10^{-13}i$ &$ -3.602  + 2.63 \times 10^{ -13}i $\\
\hline
8  & $ -4.180  + 2.47 \times 10^{ -14}i $&$   -4.199 +  3.82 \times 10^{ -14}i  $&$-4.216+4.88\times 10^{-14}i$ &$ -4.176 +  2.24 \times 10^{ -14}i $\\
\hline
9  & $ -4.755  + 2.05 \times 10^{ -15}i $& $ -4.772  +  3.09 \times 10^{ -15}i  $&$-4.794+4.03\times 10^{-15}i$ &$  -4.751 + 1.89 \times 10^{ -15}i $\\
\hline
10 & $ -5.331  + 1.76 \times 10^{ -16}i $&  $-5.346  + 2.51 \times 10^{ -16}i   $&$-5.369+3.26\times 10^{-16}i$ &$ -5.326  +  1.58 \times 10^{ -16}i $\\
\hline
11 & $-5.907   + 1.49 \times 10^{ -17}i$&   $-5.921  + 2.03 \times 10^{ -17}i   $&$-5.947+2.65 \times 10^{-17}i$  &$ -5.902  + 1.32 \times 10^{ -17}i $\\
\hline
12 &$ -6.483   + 1.22 \times 10^{ -18}i$ &  $-6.496  + 1.65 \times 10^{ -18}i   $&$-6.516+ 2.07\times 10^{-18}i $&$-6.477  + 1.09 \times 10^{ -18}i $\\
\hline
13 & $-7.059   + 1.04 \times 10^{ -19}i $&   $-7.071 + 1.34 \times 10^{ -19}i  $ &$-7.102+1.81\times 10^{-19}i $ &$ -7.053  + 8.97 \times 10^{ -20}i $\\
\hline
\end{tabular}
\end{center}
\caption{\label{tab:efficiency3} Some numerical values of the
instability for a geometry with $m=5,\, n=1,\, c_1=1.10,\,
c_5=1.52,\, a_1=262.7,\, \lambda=m_{\phi}=0$ and $l=m_{\psi}$. In
the second column, we have the results of the full numerical
analysis; in the third column, WKB$_{\rm NUM}$ (numerical WKB)
stands for values obtained using the full WKB approximation,
formulae (\ref{WKBparameters1}), (\ref{WKBparameters2}),
(\ref{ressonance freq}) and (\ref{tau}); and in the fourth column,
labelled  as WKB$_{\rm AN}$, the values obtained using the
parabolic approximation for the potential, formulae
(\ref{parabola})-(\ref{alpha 4}), are given. In the final column,
we present the results of the matching procedure
(\ref{eq:asymp2_quantization}),(\ref{deltaN}). Notice the close
agreement between all the different methods. For $m_{\psi}=l=1$
and for these particular values of the parameters, the WKB
analysis, as done here, breaks down. Indeed, for $m_{\psi}=1$, the
potential $V_+$ has no minimum.}
\end{table}

\section{\label{conclusion} Discussion}

In this paper, we have shown that the non-supersymmetric \jmart
solitons \cite{ross} are classically unstable. The relevant
instabilities are quite generic to spacetimes which have an
ergoregion but are horizon-free \cite{friedman}. However, as noted
in Section \ref{free}, the general proof does not strictly apply
to the \jmart solutions since the latter support nonradiative
negative energy modes as shown in Appendix \ref{sec:A2}. Hence we
have explicitly shown that the ergoregion instabilities are active
in the \jmart geometries using three different approaches, which
in the end show a remarkable agreement --- see Fig.
\ref{fig:fixNvaryl_compare_methods} and Table
\ref{tab:efficiency3}. Perhaps the most physically intuitive
method is the WKB analysis  carried on in Sec. \ref{wkb}. This
approach allows us to clearly identify the nature and physical
properties of the instability. However, this analysis is only
expected to be valid for large angular momentum quantum numbers,
\ie $m_\psi\gg0$, which is not where the instability is strongest.
The more unstable modes were studied using the matched asymptotic
expansion method \cite{staro1} in Sec. \ref{sec:Match}. As a final
consistency check of these analytical results, we made a numerical
analysis of the wave equation in Sec. \ref{numerical}.

In passing we note by considering orbifolds, the \jmart solutions
were extended to a six-parameter family which includes a third
integer $k$ characterizing the orbifold group $\mathbb{Z}_k$
\cite{ross}. Of course, it is straightforward to adapt our
instability analysis so that the modes respect this orbifold
symmetry in the covering space and so one concludes that the
ergoregion instability arises in these orbifold geometries as
well.

Let us now summarize some of the features of the ergoregion
instability found for the \jmart solutions:

\noindent (i) The general shape of the WKB potentials $V_\pm$ are
sketched in Fig. \ref{fig:potential} for the case in which an
instability is present. The key point is that when the ergoregion
is present the bottom of the potential well in $V_+$ reaches
negative values. The unstable modes are those whose pattern speed
$\Sigma_{\psi}$ is negative and approaches the minimum of $V_+$ from
above (see Fig. \ref{fig:potential}). Thus, they are nearly bound
states of the potential well in $V_+$ that can however tunnel out
to infinity through $V_-$.

\noindent (ii) The fact that the unstable modes are those with
negative phase velocity, $\Sigma_{\psi}<0$, has a clear physical
interpretation. As in the discussion of Eqs.~(\ref{Sigma}) and
(\ref{Omega}), modes with $\Sigma_{\psi}<0$ are those that propagate in
the same sense as geometry's rotation $\Omega_{\psi}$. Therefore
at infinity these modes carry positive angular momentum (same
sense as $\Omega_{\psi}$), as well as positive energy. Hence by
conservation of energy and angular momentum, with the onset of the
ergoregion instability, the \jmart solutions are shedding both
energy and angular momentum by an amount that increases
exponentially.

\noindent (iii) The instability can be quite strong, depending on
the particular combination of parameters that define the geometry.
More importantly, the instability is robust, in the sense that it
exists for a wide range of parameters.

\noindent (iv) With $m=n+1$, the \jmart solutions are
supersymmetric and so must be stable. It is a consistency check of
our analysis then that we find no instability in this case. As
commented in  Section \ref{wkb}, when $m=n+1$ the potential $V_+$,
as given by Eq.~(\ref{potential SUSY}), is always positive. Hence
there are no negative $\Sigma_{\psi}$ modes which could intersect the
potential well of $V_+$ and the SUSY geometry is stable as
required.

In our analysis, we have focused on the special case $\lambda=0$ and
$m_\phi=0$, to simplify the relevant equations. In fact, the
ergoregion instability persists when either or both of these
parameters are nonvanishing. A discussion of the general situation
is given in Appendix \ref{beyond}. The result is most simply
understood from the point of view of the WKB approach. Then all of
the additional contributions to the effective potential (\ref{rad
eq01}) introduced by a nonvanishing $m_\phi$ or $\lambda$ are
suppressed by inverse powers of $m_\psi$ and so can certainly be
neglected in the limit of large $m_\psi$. One can further check that
the instability exists over some range even when $m_\psi$ does not
dominate the other two. One distinguishing feature of $\lambda\ne0$
is that asymptotically the scalar modes have an effective mass in
five dimensions. In our analysis, this is reflected in the fact that
asymptotically $V_\pm\rightarrow\pm|\lambda/m_\psi|$ and so there is
an additional barrier for the modes to tunnel out to infinity.
However, for sufficiently large $m_\psi$, such tunnelling is
possible. One other interesting point about the large $m_\psi$
regime is that unstable modes appear with either sign of $m_\phi$
and $\lambda$. Hence while for the modes on which we have focussed,
the instability is `powered' by $J_\psi$ and results in decreasing
this angular momentum, there are unstable modes which may at the
same time increase $|J_\phi|$ and/or $P$.

Adding a mass for the scalar field modifies the potentials $V_\pm$
in essentially the same way as having nonvanishing $\lambda$. Hence
we expect the ergoregion instability will even appear for massive
fields, at least in modes with sufficiently large angular momentum.
As described in section \ref{free}, the arguments given by Friedmann
\cite{friedman} are quite general and so we expect the ergoregion
instability to appear for higher spin fields as well. In particular,
we expect the fields of the low energy type IIb supergravity will
generically experience this instability.
Having said that the ergoregion instability is robust, we must
also add that it can be suppressed in certain parameter regions.
In particular, one finds that the instability timescale becomes
extremely long in the regime where $Q_1$ and $Q_5$ are much larger
than the other scales. Further we add that in the decoupling limit
where one isolates an asymptotically AdS$_3$ core \cite{ross}, the
ergoregion instability is absent. The simplest way to understand
this result is that the AdS$_3$ core has a globally timelike
Killing vector \cite{ross} and so there is a `rotating' frame
where we can define all energies to be positive. One can also
explicitly verify the absence of an ergoregion instability in the
core solutions by directly applying the analysis used in this
paper to those backgrounds.

\begin{figure}
\centerline{\includegraphics[width=8 cm,height=6 cm]
{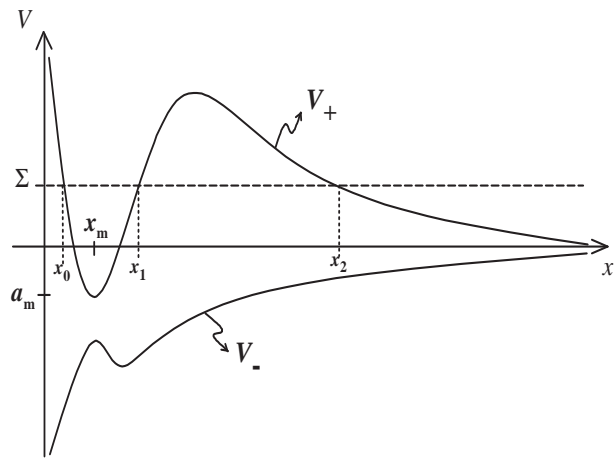}} \caption{ Damped modes are those that have
positive $\Sigma_{\psi}$. } \label{fig:potential damped}
\end{figure}

The \jmart geometries (both supersymmetric and non-supersymmetric)
also have damped modes, \ie modes (\ref{separation ansatz}) for
which the imaginary part of $\omega$ is negative. As per the WKB
analysis, these are modes with positive $\Sigma_{\psi}$ below the local
maximum of $V_+$ that tunnel out to infinity through $V_+$ --- see
Fig. \ref{fig:potential damped}.

\begin{figure}
\centerline{\includegraphics[width=8 cm,height=6 cm]
{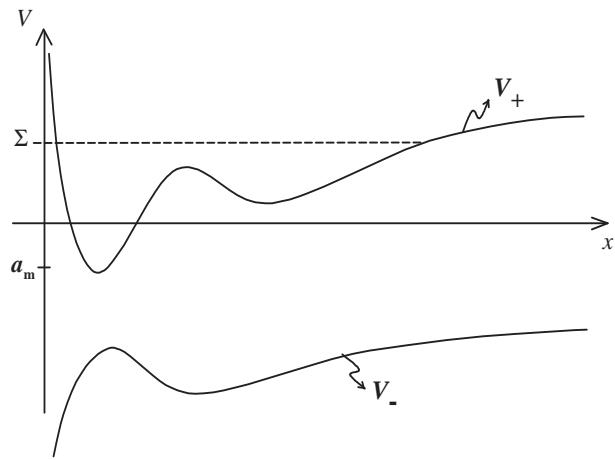}} \caption{ Qualitative shape of the potentials
$V_+$ and $V_-$ when $\omega^2-\lambda^2<0$. These are the purely
bound states that are discussed in Appendix \ref{sec:A2}. }
 \label{fig:potential bound}
\end{figure}

As emphasized previously, we can also find purely bound states
(\ie nonradiative modes) with
$\kappa^2\propto\omega^2-\lambda^2<0$. With some fine-tuning, it
may also be possible to find geometries which support bound states
with $\kappa=0$. These nonradiative modes are described in
Appendix \ref{sec:A2}. The typical situation for such modes is
sketched in Fig. \ref{fig:potential bound}. As already noted above
when $\lambda\ne0$, asymptotically
$V_\pm\rightarrow\pm|\lambda/m_\psi|$ and so there is a finite
potential barrier at infinity. If this barrier is sufficiently
large relative to $\Sigma_{\psi}=\omega/m_\psi$, bound states can
arise. These bound states can also be negative energy states, as
can be seen with the energy integral (\ref{canon}). The absence of
such negative energy modes which do not radiate at infinity was
central to Friedman's general argument for the ergoregion
instability. In \cite{friedman}, he did not find any such
nonradiative modes because he only considered the massless fields
for which there is no potential barrier at infinity. Note, however
that the current situation is more complicated because the
KK-momentum of the background, as well as the angular momenta,
contribute to the presence of an ergoregion.

The appearance of negative energy states in the presence of an
ergoregion can be anticipated from a geodesic analysis
\cite{cominsschutz}. By definition, the Killing vector $t^a$,
which generates asymptotic time translations, becomes space-like
inside the ergosphere. Hence (time-like or null) geodesics can
have either positive or negative energy, $e=-t\cdot u$, in this
region. However, asymptotically only positive energy (\ie
future-oriented) geodesics are physical. Therefore any negative
energy geodesics must be confined to circulate within the
ergoregion. Of course, in a black hole background, such geodesics
would `disappear' behind the event horizon. However, for
horizon-free geometries, such as the \jmart solutions, they are
stable bound orbits and so it is natural to find bound states in
the context of a field theory analysis. However, the question then
becomes whether  the analogous modes of the field `fit' inside the
ergoregion or whether they `leak' out to infinity, \ie whether a
negative energy bound state or an ergoregion instability results.
A more thorough examination of the bound states shows that the
negative energy bound states states are characterized by having
$\Sigma_y=\omega/\lambda<0$ while the ergoregion instability is
associated with modes where $\Sigma_{\psi}=\omega/m_\psi$ and/or
$\Sigma_{\phi}=\omega/m_\phi$ are negative -- see Appendices
\ref{beyond} and \ref{sec:A2}. Hence as the geodesic analysis
would suggest the negative energy modes have a negative pattern
speed or phase velocity, but the KK-momentum modes tend to lead to
bound states while the spinning modes are related to
instabilities.

The presence of negative energy bound states can also be expected
to enhance the decay of these horizon-free geometries. The
analysis of the ergoregion instability (considered in this paper)
is only at the level of linearized test fields. Generically any
theory coupling to gravity will also have nonlinear interactions
(\eg even the free scalar considered here has nonlinear couplings
with gravitons). These nonlinear couplings might be expected to
lead to processes, where positive energy modes are radiated at
infinity while negative energy modes are populated within the
ergoregion. However, one should note that the negative energy
modes are exponentially decaying at large radius --- see Appendix
\ref{sec:A2} --- while the positive energy modes are power-law
suppressed inside the ergoregion. Hence the overlap of these modes
is expected to be small, which will suppress this nonlinear
contribution to the decay.

We now turn to consider the endpoint of the ergoregion
instabilities. As emphasized before, the presence of these
instabilities relies on two key ingredients, namely, the geometry
has an ergoregion but it does not have an event horizon. Hence the
resulting decay process could be terminated either by the
disappearance of the ergoregion or the appearance of a horizon.
However, the unstable modes radiate with a positive energy density
asymptotically which is compensated for by a negative energy
density inside the ergoregion --- as could be seen in
Eq.~(\ref{canon}). Hence the onset of the ergoregion instability
produces a(n exponential) build-up of negative energy near the
core of the \jmart solutions. Therefore it seems unlikely that an
event horizon will form since the latter is typically associated
with a large build-up of positive energy density. This reasoning
then suggests that the decay must terminate with the disappearance
of the ergoregion. The supersymmetric D1-D5-P microstate
geometries \cite{two,three1,three12,three2} are all free of an
ergoregion and hence it is natural to suppose that these are at
the endpoint of the ergoregion instabilities. Of course, these
solutions offer a huge family of possible endpoints and the
precise one that forms will depend on the details of the decay
process, beyond the linear regime considered here --- although as
we are only considering the classical evolution, it is in
principle possible given a certain set of initial conditions. Of
course, we can expect that the final mass should be close to the
BPS mass determined by the charges of the initial \jmart solution,
\ie $E=\pi/4G_5\,[Q_1+Q_5+Q_P]$. Although even here, we can only
say `close' as we know that the unstable modes with $\lambda\ne0$
(and {\it either} sign of $\lambda$) occur which may modify the
final value of $Q_P$. Similar comments apply for the angular
momenta, $J_\psi$ and $J_\phi$. We also observe that there is no
reason to expect that the decay process will lead to an endpoint
within the family of supersymmetric \jmart solutions. Of course,
at the level of the present discussion, we cannot rule out that
the endpoint is only a nearly supersymmetric solution (or that
this would be the effective endpoint). Our expectation is that
such solutions will have a `small' ergoregion and that the
instability might be eliminated (or strongly suppressed) before
the ergoregion precisely vanishes.\footnote{We should note that
the \jmart solutions begin in a low-mass regime where
$M^2<(a_1-a_2)^2$, however, if the ergoregion instability sheds
the background angular momentum efficiently then the system will
evolve to a regime where black holes can form. Hence we can not
rule out the appearance of an event horizon -- we thank Simon Ross
for correspondence on this point.}

The stability analysis of the \jmart solitons \cite{ross} is
relevant for the stringy tachyon decays discussed recently in
\cite{rossTC}. Originally, \cite{horowitzTC} considered tachyon
condensation in certain D1-D5 black string backgrounds where
tachyonic string winding modes can occur if one chooses
antiperiodic boundary conditions for the fermions around the
circle on which the black string is compactified. The latter
choice necessarily restricts the scenario to a non-supersymmetric
sector of string theory which already suffers from various
instabilities \cite{negative}. Ref. \cite{rossTC} considered
adding angular momentum to the black strings. In this case, it was
shown that string modes winding certain compact circles near the
horizon can be tachyonic even when the asymptotic fermion boundary
conditions are supersymmetric. The relevant point for the present
discussion is that the endpoint of the tachyon condensation is in
general one of the non-supersymmetric \jmart solitons. Now, in
this paper, we have shown that these solitons are themselves
unstable and so they will not be the final endpoint of these
decays. Instead, the ergoregion instability will continue the
decay process and as suggested above, will likely terminate with a
supersymmetric microstate geometry.

We would now like to consider the implications of ergoregion
instabilities for Mathur's fuzzball program of describing black
holes in terms of an ensemble of microstate geometries. If this
program is to succeed it must supply a description of both
supersymmetric and also non-supersymmetric black holes. At first
sight, it may appear that constructing non-BPS microstate geometries
is not possible. In particular, non-BPS states will decay and so it
is not clear that there should be stationary geometries to describe
them. However, the \jmart solutions provide an explicit example
indicating that this is not really a problem. In fact, the decay of
non-BPS microstates was already considered in the D-brane
description of nonextremal black holes \cite{revall}. In that
context, it was seen as a success of the string theoretic approach
as this instability had an interpretation in terms of Hawking
radiation \cite{curt,malstrom,sdsm}. Of course, Hawking radiation is
a quantum effect in the black hole background and so presents no
obstacle to the construction of classical supergravity solutions
which are static or stationary.

It is perhaps useful to remind ourselves as to how this
distinction arises. The classical limit can be understood as the
limit where the string coupling $g_s$ is vanishingly small
\cite{malstrom}. However, the interesting classical solutions are
those which correspond to states where the various quantum numbers
are extremely large. That is, $n_1,n_5\propto 1/g_s$ and
$n_p,J_\psi,J_\phi\propto1/g_s^2$ while $g_s\rightarrow0$. These
scalings are chosen to ensure that the gravitational `footprint',
\ie $Q_1$, $Q_5$, $Q_P$, $a_1$ and $a_2$, associated with each of
these quantum numbers remains finite. However, in this limit, the
ADM energy of the system diverges with $E\propto 1/g_s^2$. As the
energy is a dimensionful quantity, this can be accommodated by
changing the units with which energies are measured in the
classical limit. Essentially this divergence is associated with
the divergence of the Planck mass, which does not serve as a
useful reference scale in classical gravity. Now the decay rate of
the nonextremal D1-D5-P black holes can be computed in a
straightforward manner \cite{malstrom,sdsm}. The key point,
however, is that the final expression for $dE/dt$ is expressed in
terms of geometric quantities and is independent of $g_s$.
Therefore in the classical limit, the rate of energy loss remains
finite in stringy units but becomes vanishingly small when
measured against the fiducial energy scale that was established
for classical physics. Hence the non-BPS black holes become stable
in the classical regime.

We note that the `straightforward' calculations of the decay rate
referred to above can be performed either in the framework of a
microscopic D-brane perspective or of the gravitational perspective
of Hawking radiation. The suprising result is that the results of
both analyses  agree precisely \cite{malstrom,sdsm}, including
greybody factors, at least in the so-called `dilute gas'
approximation \cite{dilute}. However, even though suggestive
arguments can be made in this regime \cite{poor}, this remarkable
agreement remains poorly understood. As the \jmart solutions are
horizon-free, the gravitational calculation of the decay rate would
have to be modified. Using the connection between absorption and
emission rates, it is possible that absorption calculations along
the lines of those presented in \cite{ross} could be extended to
yield the desired decay rate. On the other hand, the underlying
microscopic states for the \jmart solutions were already identified
in \cite{ross}. Hence one can use microscopic techniques to estimate
the decay rate expected for these solutions. The result is
$dE/dt\sim Q_1Q_5(m-n)^6/R^6$ and again this quantity remains finite
as $g_s\rightarrow0$. Therefore we can again ignore this decay
channel for the classical \jmart solutions.

However, the ergoregion instability investigated in this paper is a
classical instability and so should not be associated with the decay
discussed above. We should also note that the form of these two
instabilities differs. Above one is considering the spontaneous
decay of the system while the classical instability really
corresponds to a decay that results when the initial data does not
precisely match that of the \jmart solutions. Of course, in the
quantum regime, the same modes associated with the ergoregion
instability will give rise to spontaneous decay due to quantum
fluctuations of the background.\footnote{In \cite{ross} it was
erroneously assumed that all of these geometries have an AdS$_3$
core to argue that such emissions would not occur.} However, the
latter will again be suppressed in the $g_s\rightarrow0$ limit. This
reflects the fact that the background can be prepared with
arbitrarily accurate precision in the classical limit and so it
should be possible to produce an arbitrary suppression of ergoregion
instability. Alternatively, working in the classical limit, we can
regard the ergoregion instability as a property of how the \jmart
solutions interact with external sources. That is, generically if an
external wave packet impinges on one of the non-supersymmetric
\jmart configurations, it will produce a dramatic decay of the
original background. Hence this instability seems to present a major
challenge for the fuzzball description of black holes.

We have argued that the ergoregion instability is a robust feature
of the non-supersymmetric \jmart solutions over a wide range of
parameters. Given general arguments along the lines of
\cite{friedman}, we also expect that this instability will  be a
generic feature of any smooth horizon-free geometries which describe
microstates which are non-BPS and carry significant angular momentum
(and hence have a macroscopic ergoregion). Therefore if a
nonextremal D1-D5-P black hole is to be described by a
coarse-grained ensemble fuzzball, it seems that that the classical
black holes must suffer from an analogous instability. While the
presence of an event horizon eliminates the possibility of an
explicit ergoregion instability, there are, in fact, a number of
potential instabilities which might afflict these black holes and
possibly reproduce the same physics:

\noindent {\bf a) Superradiant Instability}: Spinning nonextremal
black holes will exhibit superradiant scattering, where an
incident wave packet can be reflected with a stronger amplitude.
Superradiance by itself does not provide a classical instability,
but an instability can arise if the scattered modes are reflected
back to rescatter, as described in Section \ref{general}. This
scattering was considered for higher dimensional spinning black
branes \cite{cardoso} and there it was found that when the
noncompact space has more than four dimensions, this instability
does not arise. Explicitly analyzing the present D1-D5-P black
string again seems to indicate the absence of an instability
\cite{more2}.

\noindent {\bf b) Gyration Instability}: Considering
supersymmetric D1-D5-P black strings, it was found that above a
certain critical angular momentum a straight black string is
unstable towards carrying the angular momentum in gyrations of the
horizon \cite{donny}. This instability should also appear in
non-supersymmetric configurations and so would present an
instability at large values of the angular momentum.

\noindent {\bf c) Gregory-Laflamme Instability}: The relevant
configurations are black strings and so are expected to suffer
from the Gregory-Laflamme instability \cite{grelaf} in two ways.
The first is the usual instability of long wavelength modes along
the string. Of course, this instability can be eliminated by
reducing the radius of the compactification along the string. For
a fixed radius, it is also suppressed by the boosting along this
direction which induces the KK-momentum \cite{boost}. This
instability is not related to the angular momentum carried by the
black string or the presence of an ergoregion, but we list it here
for completeness.

\noindent {\bf d) Ultra-spin Instability}: In six or higher
spacetime dimensions, one can find black hole solutions with an
arbitrarily large spin per unit mass \cite{myersperry}. However,
it was argued \cite{ultra} that a Gregory-Laflamme-like
instability will arise to dynamically enforce a Kerr-like bound in
these cases. While this analysis does not directly apply in five
dimensions, entropy arguments suggest an analogous instability
still exists and will lead to the formation of a black ring if the
angular momentum is too large \cite{ring}.

While there are several possibilities for instabilities of a black
string in six dimensions, it seems that none of these can
reproduce the physics of the ergoregion instability which will
afflict the non-BPS microstate geometries. This observation relies
on the fact that these instabilities have a different character at
a very basic level. The ergoregion instability might be termed a
radiative instability, in that, the instability is by definition
connected to modes that radiate at infinity. In contrast, the four
instabilities considered above for black strings can be termed
internal instabilities. That is, these instabilities are primarily
associated with a rearrangement of the internal or near-horizon
structure of the black string. While these instabilities will be
accompanied with some radiation at infinity, this will be a
secondary effect with these instabilities. Therefore it seems that
emulating the ergoregion instability in a nonextremal black string
background will require the discovery of a new kind of
instability. While we are performing a detailed analysis of the
nonextremal D1-D5-P black string, our preliminary results indicate
that no such instability arises \cite{more2}.

We also note in passing that at the same time the microstate
geometries should be able to emulate any instabilities found in
the black string backgrounds. In particular, the Gregory-Laflamme
instability is a robust instability that will afflict these
backgrounds for sufficiently large $R$. In the microstate
geometries, one should then find unstable modes carrying
KK-momentum which are confined near the core of the soliton. We
have studied bound states for a test field in the \jmart
solutions, as described in Appendix \ref{sec:A2}. While the modes
we identified only arise for nonvanishing KK-momentum as desired,
they are all stable, \ie they have real frequencies. Hence they
can not serve as the analog of the Gregory-Laflamme instability in
the non-supersymmetric \jmart solutions. However, the latter would
be a gravitational instability, \ie it should not be expected to
appear as a scalar test field, and so this question requires
further investigation.

A possible reconciliation of these ideas with the fuzzball
proposal would be that the microstate geometries could provide an
accurate description of a black hole but only over a long but
finite time. In the context of the AdS$_3$/CFT$_2$ duality, some
evidence for such a picture has recently been found \cite{vjtime}.
With this new point of view, a key question is to determine the
timescale over which microstate geometries cannot be distinguished
from black holes. One suggestion \cite{vjtime} is that it should
be of the order of the recurrence time, which would be exponential
in the relevant quantum numbers. An alternative suggestion might
be that the timescale is associated with Hawking evaporation which
would involve (inverse) powers of the quantum numbers. However,
note that both of these suggestions diverge in the classical
limit. Hence the ergoregion instability found here seems to be in
conflict with both of these suggestions. While the instability
timescale is certainly very long in certain parameter regimes, it
is a classical timescale, \ie it is finite in the classical limit.
Hence our results would suggest that spinning microstate
geometries and black holes should be distinguishable on a large
but classically finite timescale.

However, one must ask how characteristic our results for the
\jmart solutions will be of generic microstate geometries. In
particular, we note that the CFT states corresponding to the
\jmart solutions are exclusively in the untwisted sector
\cite{ross,three12,cft}. On the other hand, the majority of
microstates accounting for the entropy of the black strings are
expected to be in a (maximally) twisted sector \cite{fuzzy}. From
a geometric point of view, we would observe that the \jmart
solutions have all the same Killing symmetries as the D1-D5-P
black holes, while the generic microstate geometry is expected to
have a complex nonsymmetric core. Therefore it is not unreasonable
to expect that the ergoregion instability timescales found for the
\jmart solutions will not be characteristic of the microstate
geometries that make up `most' of the black hole.

One possibility might be the generic non-BPS geometries do not
have ergoregions despite the fact that they carry angular
momentum. However, we argue that such a scenario is implausible as
follows: The fuzzball description would now require that both the
horizon and the ergosphere arise as effective surfaces in
`coarse-graining'. However, quantum fluctuations must then extend
out to the ergosphere. In particular, these fluctuations extend to
regions of the spacetime which should be causally accessible to
asymptotic observers on finite classical timescales. Hence it
seems inconsistent to say that the underlying microstate
geometries are hidden from asymptotic observers in this scenario.

Hence as argued above, if the non-BPS microstate geometries are
horizon-free with an ergoregion, they should expect an ergoregion
instability. However, it may be that instability timescales
calculated for the \jmart solutions are not representative of
those for typical microstate spacetimes. In particular, the latter
should have complicated throats --- as seen in their
supersymmetric counterparts \cite{10,19,two} --- which would
emulate the absorptive behavior of a black hole horizon. Hence it
might be expected that the relevant timescales are extremely long.
An important question is then whether the instability timescale is
classically finite or not. That is, will this timescale diverge as
the quantum numbers grow as described above. Certainly finding
more generic non-BPS microstate geometries is an essential step
towards resolving this issue.

In closing, we note that in the context of the AdS/CFT, a complete
description has been produced for half-BPS microstate geometries
with AdS$_5$ \cite{lin} and AdS$_3$ \cite{ads3} asymptotics. This
framework has given rise to an interesting program of
semi-classical quantization \cite{vjtime,semi,semi3} and a
coarse-graining description of spacetime geometry \cite{emerge}.
With this program in mind, it is useful to recall the role of the
smooth horizon-free microstate geometries in Mathur's `fuzzball'
program \cite{fuzzy}.

The BPS microstate geometries for the D1-D5 system can be derived
by studying the F1-P geometries and applying a series of duality
transformations \cite{two}. There the winding and wave numbers
might be quantized by the geometry but classically the amplitudes
of the string excitations are continuous variables. Solutions
where select modes are excited with a large amplitude can then be
seen as `coherent states' of the underlying quantum theory. Such
solutions may be further useful to understand certain properties
of typical microstates, \eg their transverse size \cite{fuzzy}.
However, ultimately a generic state will have a vast number of
modes excited with very few quanta and hence the corresponding
`spacetime' will not be accurately described by a classical
geometry. However, the family of classical geometries still serve
as a guide to the classical phase space which must be quantized
\cite{semi3}.

In the present context, we wish to go beyond the BPS sector where
program is much less developed. In particular, we still face the
challenge of constructing a more or less complete family
microstate geometries. The existence of the \jmart solutions
indicate that at least certain non-BPS states can be described by
classical geometries. However, it is not at all clear how large a
class of nonsupersymmetric smooth horizon-free geometries exists.
Going beyond the present special class of solutions will probably
call for the development of new solution-generating techniques,
but the \jmart geometries offer hope that a broader class of
nonsupersymmetric solutions can be found. This will certainly be
an intriguing direction for further research and will undoubtedly
lead to interesting new insights and discoveries.

\section*{Acknowledgements}

It is a pleasure to acknowledge Vijay Balasubramanian, Thomas Levi,
Donald Marolf, Samir Mathur and Simon Ross for interesting comments
and discussions. Research at the Perimeter Institute is supported in
part by funds from NSERC of Canada and MEDT of Ontario. RCM is
further supported by an NSERC Discovery grant. VC and OJCD
acknowledge financial support from Funda\c c\~ao para a Ci\^encia e
Tecnologia (FCT) - Portugal through grant SFRH/BPD/2004. OJCD also
acknowledges CENTRA - Centro Multidisciplinar de Astrof\'{\i}sica,
Portugal for hospitality. JLH was supported by an NSERC Canada
Graduate Scholarship.




\appendix

\section{\label{sec:A0} WKB matching}


In this appendix we use the usual WKB wavefunctions and WKB
connection formulae at the turning points to relate the amplitude
of the wavefunctions in the four distinct regions of the
scattering problem and, in particular, to derive
(\ref{connectionWKB}). The four WKB regions are (see Fig.
\ref{fig:potential}): Region I, the innermost forbidden region
($0<x<x_0$); Region II, the allowed region where $V_+$ is below
$\Sigma_{\psi}$ ($x_0<x<x_1$); Region III, the potential barrier region
where $V_+$ is above $\Sigma_{\psi}$ ($x_1<x<x_2$); and Region IV, the
external allowed region where $\Sigma_{\psi}$ is below $V_-$
($x_2<x<\infty$). The WKB wavefunctions in region I and in region
IV were already written in (\ref{H1}) and (\ref{H4}),
respectively, and in regions II and III they are given by
\begin{eqnarray}
H_{\rm II}&\simeq& \frac{C_2}{m_{\psi}^{1/2}T^{1/4}} {\rm
exp}\left[ i\,m_{\psi}\int_{x_1}^{x} \sqrt{T}\,dx \right]
+\frac{C_3}{m_{\psi}^{1/2}T^{1/4}} {\rm exp}\left[
-i\,m_{\psi}\int_{x_1}^{x} \sqrt{T}\,dx \right], \label{H2}  \\
H_{\rm III}&\simeq& \frac{C_4}{m_{\psi}^{1/2}|T|^{1/4}}
 {\rm exp}\left[ -\,m_{\psi}\int_{x_1}^{x} \sqrt{|T|}\,dx \right] +
 \frac{C_5}{m_{\psi}^{1/2}|T|^{1/4}} {\rm exp}\left[
\,m_{\psi}\int_{x_1}^{x} \sqrt{|T|}\,dx \right]. \label{H3}
\end{eqnarray}
Using the WKB connection formulae in each turning point, $x_0$,
$x_1$ and $x_2$, we can find the relations between the amplitudes
$C_i$'s $(i=1,\cdots,7)$ of the several regions, yielding:
\begin{equation}
C_2=C_1 e^{i \gamma}\,, \quad C_3=C_1 e^{-i \gamma}\,.
 \label{connectionWKB-12}
\end{equation}

\begin{equation}
C_4= \frac{1}{2} \left (C_2 e^{-i \pi/4}+C_3 e^{i \pi/4}\right)\,,
\quad C_5 =i \left (C_2 e^{-i \pi/4}-C_3 e^{i \pi/4}\right)\,,
 \label{connectionWKB-23}
\end{equation}

\begin{equation}
C_6 =  \left ( \frac{i C_4}{2\eta}+ C_5 \eta \right )e^{-i
\pi/4}\,,\quad C_7 = \left ( -\frac{i C_4}{2\eta}+ C_5 \eta \right
)e^{i\pi/4}\,,
 \label{connectionWKB-34}
\end{equation}
with $\gamma$ and $\eta$ defined in (\ref{WKBparameters1}) and
(\ref{WKBparameters2}), respectively. Finally, combining these
three sets of relations yields (\ref{connectionWKB}).


\section{\label{sec:A1}Details of numerical analysis}

\begin{figure}[h]
\begin{center}
\resizebox{10cm}{6cm}{\includegraphics{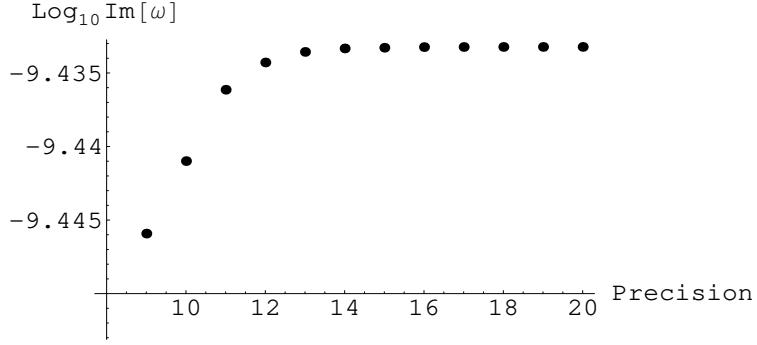}}
\end{center}
\caption{Effect of increasing digits of precision used on
imaginary part of eigenvalue.} \label{fig:compare_precision}
\end{figure}

In this Appendix we discuss some issues related to the precision
used in the numerical computations. Even though the very small
imaginary parts of $\omega$ are well described by both
approximations, for completeness we show that they are not a
numerical artifact due to loss of precision in our numeric
routines or a by-product of using the approximate solutions to
specify the boundary conditions.  In Fig.
\ref{fig:compare_precision} we plot the imaginary part of $\omega$
for several values of the number of digits of precision used in
the calculation.  We use the same parameters as before and set
$N=0$, $l=m_\psi=4$. We see, as one would expect if the imaginary
part were actually non-zero, that the eigenvalue converges to a
constant value when the number of digits is larger than the size
of the imaginary part.

\begin{figure}[h]
\begin{center}
\resizebox{7cm}{5cm}{\includegraphics{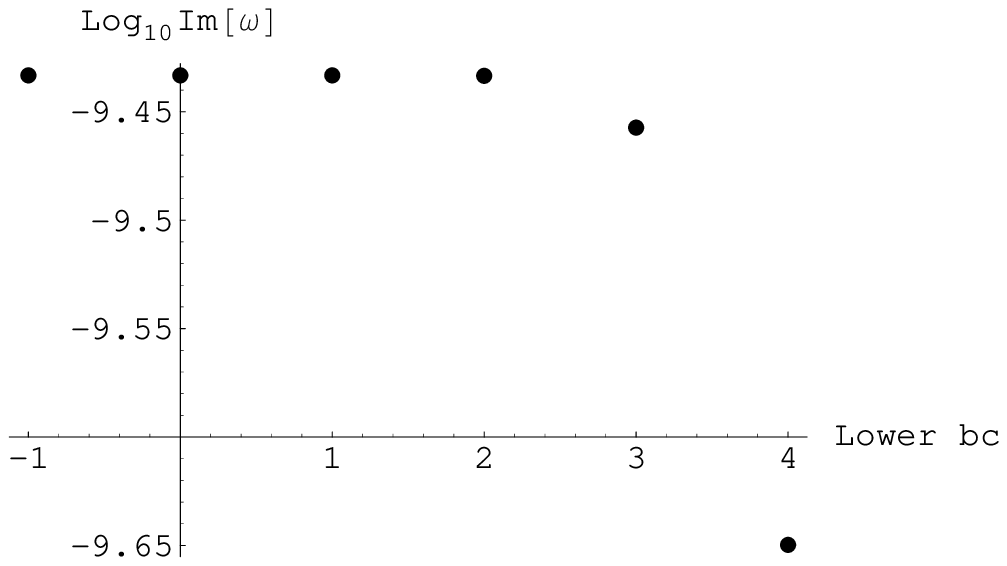}}
\resizebox{7cm}{5cm}{\includegraphics{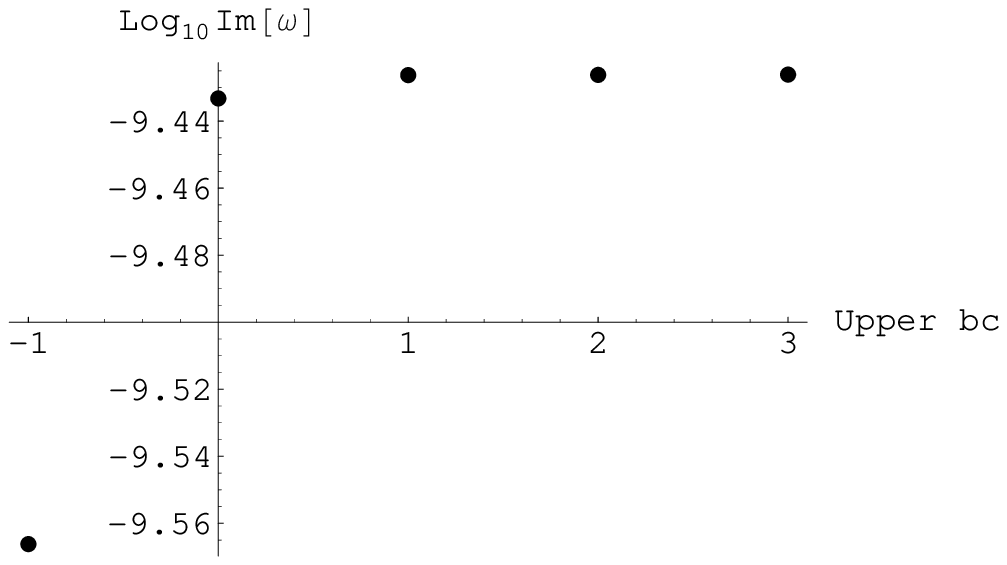}}
\end{center}
\caption{On the left we vary the point at which the inner boundary
condition is applied, $\log_{10} x_0$, from the value at which
$(x\kappa^2)/(\zeta^2/x) = 10^{-5}$.  On the right, we do the same
for the outer condition from the value at which
$(x\kappa^2)/(\zeta^2/x) = 10^{3}$.} \label{fig:compare_bcs}
\end{figure}

With only the asymptotic form of the solutions to specify the
boundary conditions we are not actually setting the coefficient on
the divergent term to zero.  Instead, there will always be some
amount of the divergent solution in the numerically defined
solution.  The suppression of the divergent term is dependent on
how deep into the asymptotic region we choose to apply the
boundary condition.  To ensure that these small divergent terms
are not causing any errors we study the effect of varying the
point at which we apply the boundary conditions.  This has been
shown in Fig.  \ref{fig:compare_bcs} on the left and right.  In
both cases, we again see that the eigenvalue converges to a
constant value as we increase the accuracy of the calculation.

\section{Detailed analysis of the instability} \label{beyond}
The existence of a solution to the matching procedure can be
reduced to the requirement that a number of constraints be
satisfied.  The difficulty one runs into when trying to discuss
the general properties of these solutions is that while all the
parameters appearing in the various equations are uniquely
determined by the set $\{Q_1,Q_5,R,m,n\}$, it is difficult to
write explicit expressions for them.  In this sense, the fact that
the parameters (\ref{simple}) can be written in such a simple form
is really quite surprising since all are proportional to $M$,
which can at best be defined implicitly in terms of the above
parameters.

Hence it is useful to have an approximation for $M$ that allows
one to understand the general behavior of the various parameters.
Surprisingly there is quite a simple approximate solution given by
\begin{equation}
M \approx 2(s^{-1}-s) \frac{Q_p}{1+nm \left ( \frac{Q_1 +
Q_5}{R^2} \right ) } \ , \label{m_approx}
\end{equation}
where we recall that $Q_p = n m Q_1 Q_5/R^2$. For most parameter
values, this expression is accurate on the order of a few percent.
When one of the D-brane charges, say $Q_1$, grows much larger than
$Q_5 \sim R^2$ this approximation can break down, though only by a
few percent times $(m-(n+1))$.  Similar problems appear when $R^2
\gg Q_1 \sim Q_5$, in this case the error appears to be of the
same order.  The important thing to note is that it gives the
correct scaling of $M$ with the various parameters in all
situations.  In most cases, except those noted previously when $m
\gg n$, it also gives the correct order of magnitude.  Treating
$m$ as a continuous parameter, the approximation appears to
produce the approach to the supersymmetric limit exactly.

Using this, one can approximate, or at least bound, the parameters
appearing in the solutions.
\begin{eqnarray}
\varrho & = & \frac{c_1^2 c_5^2 c_p^2-s_1^2 s_5^2 s_p^2}{s_1 c_1 s_5
c_5}
        \approx \frac{s^{-1}+s}{2} \left (1 + n m \frac{Q_1+Q_5}{R^2} \right ) \\
\varepsilon & \leq & \frac{1}{R^2}
        \left (Q_1 + Q_5 + Q_p + M \left ( 1-\frac{n m s^2}{(s^{-2}-s^2)^2} \right ) \right ) \nonumber \\
     & \approx & \frac{1}{R^2}
        \left (Q_1 + Q_5 + Q_p + \frac{2Q_p}{1+n m \frac{Q_1+Q_5}{R^2}}
 \left [ \frac{ nm(1-s^4)^2-s^6}{n m s(1-s^4)(1+s^2)} \right ] \right ) \\
\vartheta & \leq & \frac{Q_p}{Q_1 Q_5} \left (Q_1 + Q_5 + M \right ) \nonumber \\
 & \approx & \frac{Q_p}{Q_5} + \frac{Q_p}{Q_1}
+ 2 \frac{Q_p^2}{Q_1 Q_5} \frac{s^{-1}-s}{1+n m \frac{Q_1 +
Q_5}{R^2} } \label{background_parameters}
\end{eqnarray}
In the above, the inequalities result from writing $s_i^2 \leq s_i
c_i = Q_i/M$, in particular they become exact for the extremal
limit.  From the expression for $\varepsilon$ one sees that it is
finite, and in fact positive for all $m \geq n+2$.  It is only in
the extremal limit that $\varepsilon \rightarrow -\infty$, which
precludes any possible instability.  One may also check from these
forms that $\varepsilon /\varrho^2 \ll 1$ for all values of the
parameters, which can be verified numerically for sets of
parameters in which the approximations are less trustworthy.  In
what follows then we will neglect $\varepsilon$ where consistent.

The timescale of the instability is an increasing function of
$\nu$ which, given the above considerations, is given by
\begin{equation}
\nu \approx - \lambda \frac{Q_p}{\varrho R^2}+
        \sqrt{ \lambda \frac{Q_p}{\varrho R^2}
        \left ( \lambda \frac{Q_p}{\varrho R^2}+2 c\right )
        + \nu_0^2 - \frac{\varepsilon}{\varrho^2}c^2 } \ .
\end{equation}
Unfortunately, we cannot make any definite statements about the
size of $Q_p/\varrho R^2$ like we did previously for $\varepsilon$
since it can be made arbitrarily large or small just by varying
$R$. At this point we could use the explicit forms for $\nu_0^2$
and $c$ to discuss the general properties of the solutions.
Instead we will for now set $\lambda=0$ to make the discussion
more transparent. Non-zero $\lambda$ will not change the general
features of the solutions.

Setting $\lambda = 0$, the above expression for $\nu$ simplifies
quite a bit
\begin{equation}
\nu \approx \sqrt{\nu_0^2 - \frac{\varepsilon}{\varrho^2}c^2 } \ .
\end{equation}
We are now in a position to discuss the behavior of the timescale
for various different solutions.  Recall that the timescale for
the instability is smallest when $\nu$ is smallest.  Therefore the
instability will be strongest when $\nu_0 = l+1$ is smallest.
This, of course, does not mean that we should necessarily consider
solutions with $l=0$, in fact we shall see in a moment that such
solutions are not possible.  More precisely the minimum value of
$l$ for which all the constraints can be satisfied will lead to
the most unstable solution.

Similarly, when $c^2$ or $\varepsilon/\varrho^2$ is largest the
instability be the strongest.  We shall deal with $c$ next, but
for now it is sufficient to note that it is only dependent on $m$
and $n$. Observe from (\ref{background_parameters}) that for fixed
$R$, $\varepsilon/\varrho^2$ varies roughly like the inverse of the
charges, therefore when one considers limits in which the charges
grow, the timescale of the instability diverges.  Similar
arguments hold when $R$ is vastly different from the charges, we
find that $\varepsilon/\varrho^2$ shrinks and the lowering effect of
$c^2$ is diminished.  It appears then that the instability will be
strongest when $Q_1 \sim Q_5 \sim R^2$.


To discuss the relative effect of $c$ we should return to the
constraints.  These also simplify when we set $\lambda=0$ and we
may consider the simpler constraint $c-\nu_0 > 0$.  The exact form
that $c$ takes is dependent on the sign of $\zeta$.  By studying
the constraints, it turns out that solutions with $\zeta > 0$ will
in general exist, but for larger values of $l$ than when $\zeta <
0$. Given the considerations above, the effect of these modes will
be subdominant.  We therefore focus on $\zeta = -n m_\psi + m
m_\phi< 0$ which implies that $m_\psi > m_\phi$.  One can then
write the constraint as
\begin{eqnarray}
c-\nu_0 & = & \left [ (m-n)(m_\psi+m_\phi) - (2N+1) \right ]  -  \left [ l+1 \right ] \\
& = & (m-(n+1))(m_\psi + m_\phi) - (l-m_\psi-m_\phi) - 2(N+1) > 0
\ .
\end{eqnarray}
Further, it can be shown that when this is satisfied, the other
constraints follow automatically. The last two bracketed terms in
the final line are positive, so a solution requires that $m_\psi +
m_\phi > 0$, implying that $m_\psi$ must be positive.  This is a
general result that is also obtained when $\zeta >0$ or $\lambda
\neq 0$.  When $c$ is largest, the timescale will be shortest,
therefore the lowest harmonic $N=0$ will lead to the strongest
instability.  One can also make $c$ large by choosing $m_\psi$ and
$m_\phi$ as large as large as possible, $m_\psi+m_\phi=l$, but
taking $l$ large will not necessarily give us a very unstable mode
because as noted before it will cause $\nu_0$ to rise which has an
opposing effect.  Since $c^2$ enters weighted by
$\varepsilon/\varrho^2$, the more important contribution will be that
from $\nu_0$ and the net effect is a less unstable mode.  Finally,
note that $m_\phi$ and $m_\psi$ appear symmetrically in $c$, so
that the value of $\nu$ will be independent of the partition of
$l$ into $m_\phi$ and $m_\psi$.  This does not mean that the
timescale will be independent of this partition since it is a
weakly shrinking function of $|\zeta|$ for fixed $\nu$.  When
$|\zeta|$ is maximized the timescale will be the shortest, which
is the case when $m_\psi=l$, $m_\phi=0$.

To summarize then, for a fixed mode that solves the constraints,
the instability will be strongest when $Q_1 \sim Q_5 \sim R^2$.
On the other hand, when we fix a particular background, the
instability with $\lambda=0$ will be strongest when $l=m_\psi$ is
as small as possible and $m_\phi=0$.

Finally then we may discuss the solutions for which $\lambda \neq
0$.  It turns out that the various scalings of the other
parameters appears not to be changed.  When $\lambda \neq 0$, the
constraint $c^2 - \nu_0^2 > 0$ becomes easier to satisfy since $c$
picks up a contribution proportional to $\vartheta \lambda$ while
the contribution to $\nu_0$ is smaller.  The tougher constraint to
satisfy is then the one that implies $\omega^2 > \lambda^2$.  When
all other parameters are fixed, this places upper and lower bounds
on (\ie we allow negative $\lambda$) $\lambda$.  We will not go
into detail here, but instead note one can always find solutions
with non-zero $\lambda$ by going to sufficiently large angular
momentum, $l$.

When studying the characteristic time for the instabilities, one
finds that the timescale decreases as $\lambda$ is raised, but
reaches a maximum shortly before reaching the upper bound.  For
negative values, on the other hand, the timescale is a constant
decreasing function of $\lambda$.  As mentioned, solutions with
non-zero $\lambda$ require larger values of $l$ than when
$\lambda=0$.  Though larger $l$ tends to increase the timescale,
the overall effect of going to larger $l$ to accommodate non-zero
$\lambda$ can still lead to shorter timescales.

\section{\label{sec:A2}Bound States}

The general radial dependence of the scalar field at large
distances from the core is determined by the sign of $\kappa^{2}$.
When it is positive, the general solution oscillates with a
power-law falloff.  This is the behavior that led to the in and
outgoing waves at infinity which we have already discussed. The
other two possibilities, where $\kappa^2$ is zero or negative, can
lead to quite different behavior.  For the former there is an
exact solution, while the latter may again be solved with a
matched expansion.
\subsection{${\bm \kappa^{2}=0}$: Marginally Bound States}
By considering the special mode with $\omega^2=\lambda^2$, both
the angular and radial equations simplify sufficiently that an
exact solution may be found.  Such a choice removes all $\omega$
dependence from the angular equation allowing it to be solved
independently.  The result is the eigenvalue equation for the
harmonics on an $S^3$.  The exact eigenvalue is $\Lambda=l(l+2)$.

For the radial equation, this choice of mode removes the $\kappa^2
x$ term; the same condition that previously led to the
simplification in the near region.  The previous solution in the
near region (\ref{hypergeometric solution}) therefore becomes the
exact solution in the entire spacetime.  This means that asymptotically
the equation has a basis of solutions in terms of $r^{-1 \pm \nu}$.  Ignoring
for now the part dependent on the KK momentum, these become $r^{l}$ and $r^{-2-l}$.
These are simply the terms one expects from a Laplace series in
four flat spatial dimensions where the angular momentum creates an effective
radial potential.

Asymptotic regularity requires that the $r^{-1 + \nu}$ component
vanish whenever $\nu > 1$, leaving a field that falls off
as $r^{-1-\nu}$.  The natural generalization
of Friedmann's analysis of ergoregion instabilities to five-dimensions
would involve studying fields that fall off as $r^{-2}$, therefore
these modes will evade that analysis as long as $\nu > 1$.  The
requirement that removes the divergent term is similar to that
for outgoing modes, except now it is an exact result
\begin{eqnarray}
\nu+|\zeta| \mp \xi & = &   -(2N+1) \ , \label{eq:mbound-state-quantized}
\end{eqnarray}
where $N$ is a non-negative integer.  Here, however, we allow for
either of the $\Gamma$ functions in the denominator to diverge in
eq.~(\ref{near field large r}), leading to both possibilities for
the sign before $\xi$. This is in contrast to the search for
unstables modes in which we could neglect one of the possibilities
since it was found to corresponded to ingoing damped modes.
Indeed, since (\ref{eq:mbound-state-quantized}) contains terms
linear in both $\lambda$ and $\omega$, one must consider both
signs above in order to be consistent with the symmetry under
flipping signs as in equation (\ref{symmetry}).

In total then we have three constraints that must be satisfied for
these modes.  The first, $\omega^2=\lambda^2$, fixes $\omega$ to
be an integer, meaning that there are no remaining continuous
parameters characterizing the scalar field.  For a general
background then it is unlikely that the remaining constraints, in
particular the one defining $N$, can be solved by a judicious
choice of the integer eigenvalues. On the other hand, fixing the
set $m_{\psi},m_{\phi}$ and $\lambda$, there may be families of
backgrounds for which these marginally bound states exist.

\subsection{${\bm \kappa^{2} < 0}$: Bound States}

The final possibility for solutions of the radial equation is
$\omega^2 < \lambda^2$, or $\kappa^2 < 0$.  As in the case where
$\kappa^2$ is positive, we are unable to find an exact solution,
though progress can be made through approximation.  In particular,
since the effect of the sign of $\kappa$ is only relevant at large
distances from the core, we need only make slight modifications to
the matched asymptotic expansion analysis presented earlier.

To begin, we factor out the sign of $\kappa^2$ by redefining
$\kappa \rightarrow i\kappa$, giving solutions that are real
valued exponentials asymptotically.  Requiring regularity
therefore leaves only the exponentially damped ``bound states'',
localized near the core region.  Explicitly, after having made the
redefinition in (\ref{far wave eq}), a convenient basis of
solutions is in terms of modified Bessel functions of the first
and second kind.
\begin{eqnarray}
h & = &  \frac{1}{\sqrt{x}} \left [ A_1 I_\nu(\kappa \,\sqrt{x}) +
A_2 K_\nu (\kappa\, \sqrt{x} ) \right ] \label{eq:hout_solution} \
.
\end{eqnarray}
The first of these diverges at large $x$ and so we require $A_1=0$
for regularity.  For now though, we leave $A_1$ arbitrary, setting
it to vanish only after we have performed the matching.

In the matching region expanding $I_\nu$ and $K_\nu$ in the
$x^{\pm \nu/2}$ basis gives
\begin{equation}
h \approx \frac{1}{\sqrt{x}} \left [
 \left ( \frac{A_1}{\Gamma(1+\nu)} +\frac{A_2 \Gamma(-\nu)}{2} \right )
\left (\frac{\sqrt{x}\,\kappa}{2} \right )^\nu
 + \frac{A_2 \Gamma(\nu)}{2} \left (\frac{\sqrt{x}\,\kappa}{2} \right )^{-\nu}
  \right ] \ . \label{eq:hout_overlap}
\end{equation}
Note that $K_\nu$ contains both of these powers of $x$ when
expanded in the overlap region.  While the contribution of the
positive power to $K_\nu$ is relatively small, we will keep this
contribution until after we perform the matching so that we may
see how the approximate solution comes about.

By construction, the solution in the near region
(\ref{hypergeometric solution}) is unaffected by the redefinition of
$\kappa$.  Immediately then we may proceed to matching the
coefficients on powers of $x$ in the overlap region. Setting $A=1$
in the near region solution, we determine the coefficients $A_1,A_2$
in the outer region
\begin{eqnarray}
\frac{A_1 (\kappa/2)^\nu}{\Gamma(1+\nu) } & = &
\frac{\Gamma(\nu)\Gamma(1+|\zeta|)}{\Gamma(\frac{1}{2}(1+\nu+|\zeta|+\xi))
\Gamma(\frac{1}{2}(1+\nu+|\zeta|-\xi))}
- \frac{\Gamma^2(-\nu)\Gamma(1+|\zeta|)(\kappa/2)^{2\nu}}{\Gamma(\nu)\Gamma(b)
\Gamma(\frac{1}{2}(1-\nu+|\zeta|-\xi))}  \ , \label{eq:c1-eq}\\
\frac{A_2\Gamma(\nu)}{2(\kappa/2)^\nu} & = &
\frac{\Gamma(-\nu)\Gamma(1+|\zeta|)}{\Gamma(b)\Gamma(\frac{1}{2}(1-\nu+|\zeta|-\xi))}
\label{eq:c2-eq}\ ,
\end{eqnarray}
As before, finding the spectrum of solutions now requires that we
find values of the free parameters for which these equations are
consistent with the boundary conditions.  In particular, we now set
$A_1=0$ and therefore ask that the right hand side of
(\ref{eq:c1-eq}) vanishes.  Again, rather than find such parameters
numerically there is an accurate approximation that comes from
noting that consistency requires $A_2$ be non-zero. This implies
that the second term in (\ref{eq:c1-eq}) must be non-zero and
therefore any solution must come from cancellation between the two
terms.  Since the second term is suppressed by the factor
$\kappa^{2\nu}$, a comparable suppression must occur in the first
term, again  requiring the divergence of a $\Gamma$ function in the
denominator.  This gives a quantization condition similar to that found
previously
\begin{equation}
\nu+|\zeta| \mp \xi \approx  -(2N+1)\ .
\label{eq:bound-state-quantized}
\end{equation}
Here again, the terms linear in $\omega$ and $\lambda$ implicit in
the above equation -- see definitions in eq.~(\ref{rad eq
parameters0}) --- imply that both possibilities are required for
consistency with the symmetry (\ref{symmetry}), though in practice
both may not lead to solutions for which $\omega^2 < \lambda^2$.

When $\nu$ is real, this appears to give solutions for $\omega$
which are purely real. Note, however, we must be careful in solving
the constraint since, given the right combination of background
charges, $\nu^2$ could become negative. For an
arbitrary frequency in this range, Eq.~(\ref{eq:c1-eq}) will be
complex so solutions where $\omega$ has both real and imaginary
parts may be possible.  Such solutions cannot be found with the sort
of perturbative expansion used in studying the outgoing modes since
now it is the real part of $\nu$ which gains a small correction,
while the imaginary part is large. We can therefore no longer
consider the behavior near the pole on the negative real axis defined by the
real part of Eq.~(\ref{eq:bound-state-quantized}). Instead, we have
resorted to searching for these solutions by solving
(\ref{eq:c1-eq}) numerically.

\begin{figure}
\centerline{\includegraphics[width=8 cm,height=6 cm]
{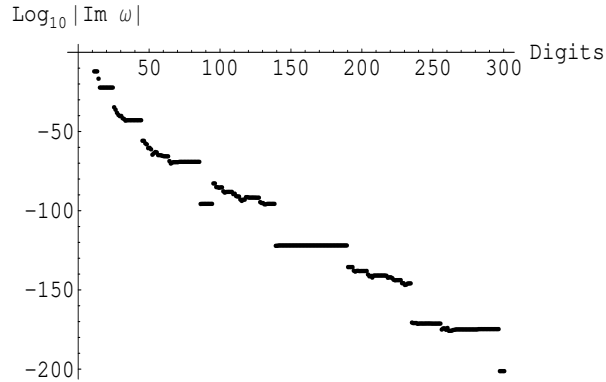}} \caption{Variation of the size of the imaginary part
of $\omega$ resulting from the numerical solution of
(\ref{eq:c1-eq}) as the precision is increased.} \label{fig:unstable_bs}
\end{figure}

Generically, the root finding algorithm will produce a complex value
of $\omega$ that sets the equation to zero within a specified
precision.  Since the imaginary part is many orders of magnitude
smaller than the real part one should ensure that it really is
non-zero and not a numerical artifact.  In Figure
\ref{fig:unstable_bs} we show the variation in the size of the
imaginary part as a function of the tolerance used in finding the
root of (\ref{eq:c1-eq}). From this plot we see that the imaginary
contribution is indeed just an artifact of trying to solve the
complex equation. Surprisingly then it appears we can satisfy
(\ref{eq:c1-eq}) with a real value of $\omega$, even if that value
causes $\nu^2 <0$. That value corresponds to the solution of the
equation resulting from taking the real part of the quantization
conditions (\ref{eq:bound-state-quantized}).

Since the condition (\ref{eq:bound-state-quantized}) is the same
as for the outgoing modes, much of the analysis in Appendix
\ref{beyond} about the existence of solutions applies.  The
situation is somewhat more complicated in that one now allows
modes with positive $\omega$ and there are two possible solutions
corresponding to the two signs in
(\ref{eq:bound-state-quantized}), but the general characteristics
of the solutions are the same.  In particular, for the outgoing
modes it was found that there are upper and lower bounds on the
allowable values of $\lambda$ beyond which $\omega^2 - \lambda^2$
changes sign.  In light of these bound states, we see that the
full space of solutions may be considered as split into distinct
regions based on the value of $\lambda$.  There is a
small-$|\lambda|$ regime, in which one finds the outgoing unstable
modes. This is surrounded, at larger values of $|\lambda|$, by a
regime where the bound states arise.

This separation of the two types of modes according to the
parameter $\lambda$ makes clear the difference in their origin. In
particular, one can always find outgoing unstable modes that do
not carry KK momentum, they need only be supplied with sufficient
angular momentum.  This is in accord with our interpretation of
these solutions as the unstable modes predicted by Friedman which
result from the existence of the ergoregion. In contrast, bound
states will always result as long as $|\lambda|$ is large enough.
This includes modes which carry no angular momentum, thus
indicating the important characteristic of these solutions is
their KK momentum and the effective five-dimensional mass it
induces.

Having established the existence of these bound states we should
question just how close to the core region they are bound.  The
solutions are damped exponentially and so have characteristic size
\begin{eqnarray}
x_{bs}^{-1} \sim \kappa^2 & = &  (\lambda^2 - \omega^2) \frac{r_+^2-r_-^2}{R^2} \ , \\
& \approx & 2(\lambda^2-\omega^2) \frac{Q_1 Q_5}{(s^{-1}+s)R^2(R^2+n m(Q_1+Q_5))} \ .
\end{eqnarray}
To arrive at the final line we have used the
approximation for $M$ (\ref{m_approx}) found in Appendix
\ref{beyond}.  The boundary of the ergoregion, on the other hand, is
given by the vanishing of the norm of the Killing vector
$\partial_t$ (\ref{ergoregionV}).  We ignore the
$a_1,a_2$ dependent contributions appearing in $f$ to give an outer bound
on the size of the ergoregion, given approximately by
$r_{er}^2 \sim M c_p^2$.   In terms of the variable $x$ this means
\begin{eqnarray}
x_{er}^{-1} \gtrsim \frac{s^{-2}-s^2}{n m(s^{-2}-s^2)^2 c_p^2-s^2} \ .
\end{eqnarray}

Whenever $Q_1$ and $Q_5$ are much smaller than $R^2$, the
size of the bound state scales as $x_{bs}^{-1} \sim Q_1 Q_5/R^4 \ll 1$.
On the other hand, for large $Q_1$ and $Q_5$ we have $x_{bs}^{-1} \sim Q_i/R^2$
where $Q_i$ is the smaller of the two.  In other words, the size of the
bound state is strongly dependent on the background.  When the charges are
large, the bound state will be mostly contained within the ergoregion, while
for small charges the exponential tail of the bound state can extend far outside.

Finally we can consider the possibility that the bound states have
negative energy, which requires a detailed analysis of the energy
integral (\ref{canon}). Examining the integrand evaluated on bound
state solutions, we see that it may become negative near where the
modulus of the scalar field peaks if the latter occurs inside the
ergoregion. Though there are bound states for arbitrarily large
values of $|\lambda|$, the total energy will not be negative for
all of these. Instead, the modes that tend to exhibit negative
energy densities (in the ergoregion) only appear for a limited
range of $|\lambda|$, which is just beyond the small-$|\lambda|$
regime discussed above. That is, for values of $\lambda$ near
where the ergoregion instability appears. When this is the case,
the maximum of the modulus of the scalar field is inside the
ergoregion and the phase velocity in the compactified direction
$\Sigma_y = \omega/\lambda$ is negative, \ie in the direction
opposite to which the background is boosted.



\begin{thebibliography}{99}

\bibitem{ross} V. Jejjala, O. Madden, S.F. Ross and G. Titchener,
``Non-supersymmetric smooth geometries and D1-D5-P bound states,"
Phys. Rev. D {\bf 71}, 1240030 (2005),  hep-th/0504181.

\bibitem{2} A. Strominger and C. Vafa, ``Microscopic Origin of the
Bekenstein-Hawking Entropy," Phys. Lett. B {\bf 379} (1996) 99,
hep-th/9601029.

\bibitem{revall} A.W.~Peet, ``The Bekenstein formula and
string theory (N-brane theory),'' Class.\ Quant.\ Grav.\  {\bf 15}
(1998) 3291, hep-th/9712253;\\
S.R.~Das and S.D.~Mathur, ``The quantum physics of black holes:
Results from string theory,'' Ann.\ Rev.\ Nucl.\ Part.\ Sci.\  {\bf
50} (2000) 153, gr-qc/0105063;\\
J.R.~David, G.~Mandal and S.R.~Wadia, ``Microscopic formulation of
black holes in string theory,'' Phys.\ Rept.\  {\bf 369} (2002) 549,
hep-th/0203048.

\bibitem{big} J. M. Maldacena, ``The large N limit of superconformal
field theories and supergravity," Adv. Theor. Math. Phys. {\bf 2}
(1998) 231, hep-th/9711200;\\
E. Witten, ``Anti-de Sitter space and holography," Adv. Theor. Math.
Phys. {\bf 2} (1998) 253, hep-th/9802150.

\bibitem{adscft} O.~Aharony, S.S.~Gubser, J.M.~Maldacena,
H.~Ooguri and Y.~Oz, ``Large N field theories, string theory and
gravity,'' Phys.\ Rept.\  {\bf 323} (2000) 183, hep-th/9905111.

\bibitem{thermal1} E.~Witten, ``Anti-de Sitter space, thermal
phase transition, and confinement in gauge theories,'' Adv.\
Theor.\ Math.\ Phys.\  {\bf 2} (1998) 505, hep-th/9803131.

\bibitem{thermal2} A.~Chamblin, R.~Emparan, C.V.~Johnson and R.C.~Myers,
``Holography, thermodynamics and fluctuations of charged AdS black
holes,'' Phys.\ Rev.\ D {\bf 60} (1999) 104026, hep-th/9904197.

\bibitem{fairy} R.~Dijkgraaf, J.M.~Maldacena, G.W.~Moore
and E.P.~Verlinde, ``A black hole farey tail,'' hep-th/0005003.

\bibitem{pure} R.C. Myers, ``Pure states don't wear
black," Gen. Rel. Grav. {\bf 29} (1997) 1217, gr-qc/9705065.

\bibitem{amati} D.~Amati, ``Black holes, string theory
and quantum coherence,'' Phys.\ Lett.\ B {\bf 454} (1999) 203,
hep-th/9706157.

\bibitem{fuzzy} S.D.~Mathur, ``The fuzzball proposal for black holes:
An elementary review,'' Fortsch.\ Phys.\  {\bf 53} (2005) 793,
hep-th/0502050;\\
S.D.~Mathur, ``The quantum structure of black holes,''
hep-th/0510180.

\bibitem{11} O. Lunin and S.D. Mathur, ``Statistical interpretation of
Bekenstein entropy for systems with a stretched horizon," Phys.
Rev. Lett. {\bf 88} (2002) 211303, hep-th/0202072.

\bibitem{10} O. Lunin and S.D. Mathur, ``AdS/CFT duality and the black
hole information paradox," Nucl. Phys. B{\bf 623} (2002) 342,
hep-th/0109154.

\bibitem{19} O. Lunin and S.D. Mathur, ``The slowly rotating near extremal
D1-D5 system as a `hot tube'," Nucl. Phys. B {\bf 615} (2001) 285,
hep-th/0107113.

\bibitem{two} V. Balasubramanian, J. de Boer, E. Keski-Vakkuri, and S. F.
Ross, ``Supersymmetric conical defects: Towards a string theoretic
description of black hole formation," Phys. Rev. D {\bf 64} (2001)
064011, hep-th/0011217;\\
J. M. Maldacena and L. Maoz, ``De-singularization by rotation," JHEP
{\bf 12} (2002) 055, hep-th/0012025;\\
O. Lunin and S. D. Mathur, ``Metric of the multiply wound rotating
string," Nucl. Phys. B {\bf 610} (2001) 49, hep-th/0105136;\\
O. Lunin, J. Maldacena, and L. Maoz, ``Gravity solutions for the
D1-D5 system with angular momentum," hep-th/0212210;\\
M.~Taylor, ``General 2 charge geometries,'' hep-th/0507223.

\bibitem{early} A.A.~Tseytlin, ``Extreme dyonic black
holes in string theory,'' Mod.\ Phys.\ Lett.\ A {\bf 11} (1996) 689,
hep-th/9601177;\\
J.C.~Breckenridge, R.C.~Myers, A.W.~Peet and C.~Vafa, ``D-branes
and spinning black holes,'' Phys.\ Lett.\ B {\bf 391} (1997) 93,
hep-th/9602065.
\bibitem{early1} J.C.~Breckenridge, D.A.~Lowe, R.C.~Myers, A.W.~Peet,
A.~Strominger and C.~Vafa, ``Macroscopic and Microscopic Entropy
of Near-Extremal Spinning Black Holes,'' Phys.\ Lett.\ B {\bf 381}
(1996) 423,
hep-th/9603078;\\
M.~Cvetic and D.~Youm, ``General Rotating Five Dimensional Black
Holes of Toroidally Compactified Heterotic String,'' Nucl.\ Phys.\ B
{\bf 476} (1996) 118, hep-th/9603100.

\bibitem{three1}  O.~Lunin,
``Adding momentum to D1-D5 system,'' JHEP {\bf 0404} (2004) 054,
hep-th/0404006.

\bibitem{three12} S. Giusto, S. D. Mathur, and A. Saxena, ``Dual geometries for a
set of 3-charge microstates," Nucl. Phys. B {\bf 701} (2004)
357-379, hep-th/0405017.

\bibitem{three2} S. Giusto, S. D. Mathur, and A. Saxena, ``3-charge geometries and
their CFT duals," Nucl. Phys. B {\bf 710} (2005) 425-463,
hep-th/0406103;\\
I.~Bena and N.P.~Warner, ``One ring to rule them all ... and in
the darkness bind them?,'' hep-th/0408106;\\
I.~Bena and N.P.~Warner, ``Bubbling supertubes and foaming black
holes,'' hep-th/0505166;\\
P.~Berglund, E.G.~Gimon and T.S.~Levi, ``Supergravity microstates
for BPS black holes and black rings,'' hep-th/0505167.

\bibitem{four} I.~Bena and P.~Kraus, ``Microstates of the D1-D5-KK system,''
Phys.\ Rev.\ D {\bf 72} (2005) 025007, hep-th/0503053;\\
I.~Bena, P.~Kraus and N.P.~Warner, ``Black rings in Taub-NUT,''
Phys.\ Rev.\ D {\bf 72} (2005) 084019, hep-th/0504142;\\
H.~Elvang, R.~Emparan, D.~Mateos and H.S.~Reall, ``Supersymmetric
4D rotating black holes from 5D black rings,'' JHEP {\bf 0508}
(2005) 042, hep-th/0504125;\\
A.~Saxena, G.~Potvin, S.~Giusto and A.W.~Peet, ``Smooth geometries
with four charges in four dimensions,'' hep-th/0509214.

\bibitem{press} W.H. Press and S.A. Teukolsky, ``Floating
Orbits, super-radiant scattering and the black-hole bomb," Nature
{\bf 238} (1972) 211.

\bibitem{dnr} T.~Damour, N.~Deruelle and R.~Ruffini, ``On
Quantum Resonances In Stationary Geometries,'' Lett.\ Nuovo Cim.\
{\bf 15} (1976) 257.

\bibitem{bhb} V.~Cardoso, O.J.C.~Dias, J.P.S.~Lemos and
S.~Yoshida, ``The black hole bomb and superradiant
instabilities,'' Phys.\ Rev.\ D {\bf 70} (2004) 044039,
hep-th/0404096 [Erratum-ibid.\ D {\bf 70} (2004) 049903].

\bibitem{zel} Ya.B. Zel'dovich,
Pis'ma Zh. Eksp. Teor. Fiz. {\bf 14}, 270 (1971) [JETP Lett. {\bf
14}, 180 (1971)];\\
Ya.B. Zel'dovich, ``Amplification of cylindrical electromagnetic
waves reflected from a rotating body," Zh. Eksp. Teor. Fiz {\bf
62}, 2076 (1972) [Sov. Phys. JETP {\bf 35}, 1085 (1972)].

\bibitem{friedman}
J.L. Friedman, ``Ergosphere instability," Commun. Math. Phys. {\bf
63}, 243 (1978).

\bibitem{cominsschutz}
N. Comins and B.F. Schutz, ``On the ergoregion instability," Proc.
R. Soc. Lond. A {\bf 364} (1978) 211.

\bibitem{compute} S. Yoshida and E. Eriguchi, ``Ergoregion
instability revisited -- a new and general method for numerical
analysis of stability," MNRAS {\bf 282} (1996) 580.

\bibitem{rossTC} S.F.~Ross,
``Winding tachyons in asymptotically supersymmetric black
strings,'' hep-th/0509066.

\bibitem{horowitzTC} G.T.~Horowitz,
``Tachyon condensation and black strings,'' JHEP {\bf 0508} (2005)
091, hep-th/0506166.

\bibitem{detweiler} S.~Detweiler,
``Klein-Gordon Equation And Rotating Black Holes,'' Phys.\ Rev.\ D
{\bf 22} (1980) 2323.

\bibitem{bhbAdS} V.~Cardoso and O.~J.~C.~Dias,
``Small Kerr-anti-de Sitter black holes are unstable,'' Phys.\
Rev.\ D {\bf 70} (2004) 084011, hep-th/0405006.

\bibitem{cardoso} V.~Cardoso and S.~Yoshida,
``Superradiant instabilities of rotating black branes and
strings,'' JHEP {\bf 0507} (2005) 009, hep-th/0502206.

\bibitem{finnx} M.~Cvetic and F.~Larsen,
``General rotating black holes in string theory: Greybody factors
and  event horizons,'' Phys.\ Rev.\ D {\bf 56} (1997) 4994,
hep-th/9705192.

\bibitem{teukolskyswsh} Four-dimensional spin-weighted spheroidal harmonics
were introduced in:\\
S.A.~Teukolsky, ``Perturbations Of A Rotating Black Hole. 1.
Fundamental Equations For Gravitational, Electromagnetic, And
Neutrino Field Perturbations,'' Astrophys.\
J.\  {\bf 185} (1973) 635.\\
The generalization to higher dimensions can be found in:\\
V.P.~Frolov and D.~Stojkovic, ``Quantum radiation from a
5-dimensional rotating black hole,'' Phys.\ Rev.\ D {\bf 67}
(2003) 084004, gr-qc/0211055;\\
V.~P.~Frolov and D.~Stojkovic,
``Particle and light motion in a space-time of a five-dimensional rotating
  black hole,''  Phys.\ Rev.\ D {\bf 68}, 064011 (2003), gr-qc/0301016; \\
M.~Vasudevan, K.~A.~Stevens and D.~N.~Page, ``Particle motion and
scalar field propagation in Myers-Perry black hole spacetimes in
all dimensions,'' Class.\ Quant.\ Grav.\  {\bf 22} (2005) 1469,
gr-qc/0407030.

\bibitem{myersperry} R.C.~Myers and M.J.~Perry,
``Black Holes In Higher Dimensional Space-Times,'' Annals Phys.\
{\bf 172} (1986) 304.

\bibitem{staro1} A.A. Starobinsky,
Sov. Phys. JETP {\bf 37} (1973) 28;\\
A.A. Starobinsky and S.M. Churilov, ``Amplification of
electromagnetic and gravitational waves scattered by a rotating
black hole," Sov. Phys. JETP {\bf
38} (1973) 1;\\
W. G. Unruh, ``Absorption cross-section of small black holes,"
Phys. Rev. D {\bf 14}  (1976) 3251.

\bibitem{CardDiasYosh} V.~Cardoso, O.J.C.~Dias and S.~Yoshida,
``Perturbations and absorption cross-section of infinite-radius
black rings,'' Phys.\ Rev.\ D {\bf 72} (2005) 024025,
hep-th/0505209.

\bibitem{abramowitz} M. Abramowitz and
A. Stegun, {\it Handbook of mathematical functions}, (Dover
Publications, New York, 1970).

\bibitem{mathematica}
Wolfram Research, Inc., Mathematica, Version 5.0, Champaign, IL
(2003).

\bibitem{negative}
E.~Witten, ``Instability Of The Kaluza-Klein Vacuum,''
  Nucl.\ Phys.\ B {\bf 195} (1982) 481;\\
D.~Brill and G.T.~Horowitz, ``Negative energy in string theory,''
Phys.\ Lett.\ B {\bf 262} (1991) 437;\\
D.~Brill and H.~Pfister, ``States Of Negative Total Energy In
Kaluza-Klein Theory,'' Phys.\ Lett.\ B {\bf 228} (1989) 359.

\bibitem{curt}   C.G.~Callan and J.M.~Maldacena,
``D-brane Approach to Black Hole Quantum Mechanics,'' Nucl.\ Phys.\
B {\bf 472} (1996) 591, hep-th/9602043.

\bibitem{malstrom} J.M.~Maldacena and A.~Strominger,
``Black hole greybody factors and D-brane spectroscopy,'' Phys.\
Rev.\ D {\bf 55} (1997) 861, hep-th/9609026.

\bibitem{sdsm}
S.R.~Das and S.D.~Mathur, ``Comparing decay rates for black holes
and D-branes,'' Nucl.\ Phys.\ B {\bf 478} (1996) 561,
hep-th/9606185;\\
S.R.~Das and S.D.~Mathur, ``Interactions involving D-branes,''
Nucl.\ Phys.\ B {\bf 482} (1996) 153, hep-th/9607149.

\bibitem{dilute}
G.T.~Horowitz and A.~Strominger, ``Counting States of
Near-Extremal Black Holes,'' Phys.\ Rev.\ Lett.\  {\bf 77} (1996)
2368, hep-th/9602051;\\
J.C.~Breckenridge, D.A.~Lowe, R.C.~Myers, A.W.~Peet, A.~Strominger
and C.~Vafa, ``Macroscopic and Microscopic Entropy of
Near-Extremal Spinning Black Holes,'' Phys.\ Lett.\ B {\bf 381}
(1996) 423, hep-th/9603078.

\bibitem{poor} J.M.~Maldacena,
``D-branes and near extremal black holes at low energies,'' Phys.\
Rev.\ D {\bf 55} (1997) 7645, hep-th/9611125.

\bibitem{more2} V. Cardoso, O.J.C. Dias, J.L. Hovdebo and
R.C. Myers, in preparation.

\bibitem{donny} D.~Marolf and B.C.~Palmer,
``Gyrating strings: A new instability of black strings?,'' Phys.\
Rev.\ D {\bf 70} (2004) 084045, hep-th/0404139.

\bibitem{grelaf} R.~Gregory and R.~Laflamme,
``Black strings and p-branes are unstable,'' Phys.\ Rev.\ Lett.\
{\bf 70} (1993) 2837, hep-th/9301052;\\
R.~Gregory and R.~Laflamme, ``The Instability of charged black
strings and p-branes,'' Nucl.\ Phys.\ B {\bf 428} (1994) 399,
hep-th/9404071.

\bibitem{boost} J.L. Hovdebo and R.C. Myers, ``Black Rings,
Boosted Strings and Gregory-Laflamme,'' in preparation.

\bibitem{ultra} R.~Emparan and R.C.~Myers,
``Instability of ultra-spinning black holes,'' JHEP {\bf 0309}
(2003) 025, hep-th/0308056.

\bibitem{ring} R.~Emparan and H.S.~Reall,
``A rotating black ring in five dimensions,'' Phys.\ Rev.\ Lett.\
{\bf 88} (2002) 101101, hep-th/0110260.

\bibitem{vjtime}
V.~Balasubramanian, P.~Kraus and M.~Shigemori, ``Massless black
holes and black rings as effective geometries of the D1-D5
system,'' Class.\ Quant.\ Grav.\  {\bf 22} (2005) 4803,
hep-th/0508110.

\bibitem{cft} O.~Lunin and S.D.~Mathur,
``Three-point functions for M(N)/S(N) orbifolds with N = 4
supersymmetry,'' Commun.\ Math.\ Phys.\  {\bf 227} (2002) 385,
hep-th/0103169.


\bibitem{lin} H.~Lin, O.~Lunin and J.~Maldacena, ``Bubbling AdS space and 1/2
BPS geometries,'' JHEP {\bf 0410} (2004) 025, hep-th/0409174;\\
S.~Corley, A.~Jevicki and S.~Ramgoolam, ``Exact correlators of
giant gravitons from dual N = 4 SYM theory,'' Adv.\ Theor.\ Math.\
Phys.\  {\bf 5} (2002) 809, hep-th/0111222. \\
D.~Berenstein, ``A toy model for the AdS/CFT correspondence,''
JHEP {\bf 0407} (2004) 018, hep-th/0403110.

\bibitem{ads3} D.~Martelli and J.F.~Morales,
``Bubbling AdS(3),'' JHEP {\bf 0502} (2005) 048, hep-th/0412136;\\
J.T.~Liu, D.~Vaman and W.Y.~Wen, ``Bubbling 1/4 BPS solutions in
type IIB and supergravity reductions on $S^n\times S^n$,''
hep-th/0412043;\\
J.T.~Liu and D.~Vaman, ``Bubbling 1/2 BPS solutions of minimal
six-dimensional supergravity,'' hep-th/0412242.

\bibitem{semi} G.~Mandal,
``Fermions from half-BPS supergravity,'' JHEP {\bf 0508} (2005)
052, hep-th/0502104;\\
L.~Grant, L.~Maoz, J.~Marsano, K.~Papadodimas and V.S.~Rychkov,
``Minisuperspace quantization of `bubbling AdS' and free fermion
droplets,'' JHEP {\bf 0508} (2005) 025, hep-th/0505079;\\
L.~Maoz and V.S.~Rychkov, ``Geometry quantization from
supergravity: The case of `bubbling AdS','' JHEP {\bf 0508} (2005)
096, hep-th/0508059.

\bibitem{semi3} A.~Jevicki and A.~Donos, ``Dynamics of chiral
primaries in AdS$_3 \times S^3 \times T^4$,'' hep-th/0512017;\\
V.S.~Rychkov, ``D1-D5 black hole microstate counting from
supergravity,'' hep-th/0512053.

\bibitem{emerge} D.~Berenstein, ``Large N BPS states and
emergent quantum gravity,'' hep-th/0507203;\\
P.G.~Shepard, ``Black hole statistics from holography,'' JHEP {\bf
0510} (2005) 072, hep-th/0507260;\\
V.~Balasubramanian, J.~de Boer, V.~Jejjala and J.~Simon, ``The
library of Babel: On the origin of gravitational thermodynamics,''
hep-th/0508023;\\
V.~Balasubramanian, V.~Jejjala and J.~Simon, ``The library of
Babel,'' hep-th/0505123;\\
P.J.~Silva, ``Rational foundation of GR in terms of statistical
mechanics in the AdS/CFT framework,'' hep-th/0508081.






\end{thebibliography}
\end{document}